\documentclass[showpacs,amssymb,aps]{revtex4}
\usepackage{amsmath}
\usepackage{graphicx}
\usepackage[latin1]{inputenc}
\begin{document}

\title{Thermal Operator Representation of Finite Temperature Graphs}

\author{F. T. Brandt$^{a}$, Ashok Das$^{b}$, Olivier Espinosa$^{c}$,
  J. Frenkel$^{a}$ and Silvana Perez$^{d}$}
\affiliation{$^{a}$ Instituto de Física, Universidade de São
Paulo, São Paulo, BRAZIL}
\affiliation{$^{b}$ Department of Physics and Astronomy,
University of Rochester,
Rochester, New York 14627-0171, USA}
\affiliation{$^{c}$ Departamento de Física, Universidad
Técnica Federico Santa María, Casilla 110-V, Valparaíso, CHILE}
\affiliation{$^{d}$ Departamento de Física, 
Universidade Federal do Pará, 
Belém, Pará 66075-110, BRAZIL}

\bigskip

\begin{abstract}
Using the mixed space representation $(t,\vec p\,)$ in the 
context of scalar field theories, we prove in a simple manner
that the Feynman graphs at finite temperature are related to
the corresponding zero temperature diagrams through a simple thermal
operator, both in the imaginary time as well as in the real time
formalisms. This result is generalized to the case when there is a
nontrivial chemical potential present. Several interesting properties of the
thermal operator are also discussed.
\end{abstract}


\maketitle

\section{Introduction}

In a series of recent 
papers \cite{Espinosa:2003af,Espinosa:2005gq,Blaizot:2004bg}, 
it was shown for theories involving
scalar as well as fermion fields that every graph in momentum space at
finite temperature in the imaginary time 
formalism \cite{kapusta:book89,lebellac:book96,das:book97}
is related to the
corresponding graph of the zero temperature Euclidean field theory
through a {\em thermal operator} which has a rather simple
form. Namely, for a scalar $N$-point amplitude (at any loop) at temperature
$T$  one has
\begin{equation}\label{1}
\int \prod_{i=1}^{I}\frac{\mathrm{d}^{3}k_{i}}{(2\pi)^{3}}\
\prod_{v=1}^{V} {(2\pi)^{3}} \delta^{(3)}_v(k,p) \, 
\gamma_{N}^{(T)} = 
\int\prod_{i=1}^{I}\frac{\mathrm{d}^{3}k_{i}}{(2\pi)^{3}}\ 
\prod_{v=1}^{V} {(2\pi)^{3}} \delta^{(3)}_v(k,p) \, 
{\cal O}^{(T)}\
\gamma_{N}^{(T=0)},
\end{equation}
where
\begin{equation}
{\cal O}^{(T)} = \prod_{i=1}^{I} \left(1 + n_{i} (1-S_{i})\right).
\end{equation}
Here $I$ characterizes the total number of internal propagators and
$V$ denotes the total number vertices in the graph (with the usual
relation for the number of loops $L=I-V+1$),  
$n_{i} = n (E_{i})$ corresponds to the thermal distribution associated with the
internal propagator carrying energy $E_{i} = \sqrt{\vec{k}_i^{\ 2}+m^2}$ 
and  $S_{i} = S (E_{i})$
is a reflection operator that changes $E_{i}\rightarrow -E_{i}$ (namely, it
gives a term with $E_{i}\rightarrow -E_{i}$). We denote the internal
and the external three momenta of a graph generically by
$(\vec{k},\vec{p}\ )$ respectively
and $\delta_{v}^{(3)} (k,p)$ enforces the appropriate
three momentum
conservation at the vertex $v$. For simplicity, we have included in 
(\ref{1}) a delta function which reflects the overall conservation of
the external three momenta.
Furthermore,
$\gamma_{N}^{(T=0)}$ represents the zero temperature graph in momentum
space obtained from the Euclidean field theory. This remarkable result
is, of course, calculationally quite useful since the worrisome sum
over the internal discrete energy values (particularly at higher
loops) has already been reduced to evaluating zero temperature energy
integrals. More than that, this allows us to study many
questions of interest at finite temperature such as Ward identities,
analyticity \cite{braaten,frenkel} more directly. The original proof
of this result in
momentum space \cite{Espinosa:2005gq,Blaizot:2004bg}  
however, is quite involved and uses regularization
procedures that obscure the origin of such a relation. Furthermore,
the proof leaves one with the feeling that such a relation is
particular to the imaginary time formalism. In this paper, we discuss
a simpler proof of the thermal operator representation both in the
imaginary time formalism
\cite{kapusta:book89,lebellac:book96,das:book97} as well as in the
real time formalisms \cite{das:book97}. 
Furthermore, we also extend this relation to the case when
there is a nontrivial chemical potential \cite{lebellac:book96} and
point out various interesting aspects of this relation.

At finite temperature, it is already noted that simplifications arise
when one works not in the energy-momentum space, but rather in a mixed space
where energy has been Fourier 
transformed \cite{das:book97,Bedaque:1993fa,Das:1997gg}.
We exploit this feature to
give a simpler derivation of the thermal operator representation in
both imaginary time and the real time formalisms with and without a
chemical potential. In this paper, we will discuss in detail theories
involving scalar fields only to bring out the essential underlying
features. The
remarkable feature of the thermal operator relation, in such a
representation, is that while the finite temperature graph depends on
\begin{equation}
\gamma_{N}^{(T)} = \gamma_{N}^{(T)} (T, E_{i},\tau_{\alpha}),
\end{equation}
where $\tau_{\alpha}, \alpha=1,2,\cdots ,N$ represent the external
time coordinates of the graph, the
thermal operator depends only on $(T,E_{i})$ but not on the external
time coordinates (the zero temperature graph depends on
$(E_{i},\tau_{\alpha})$ but 
not on $T$). As a result, the time derivative operator (with respect
to external times) commutes with the thermal operator and the
discussion of our results holds equally well for theories with
fermions as well as Yang-Mills fields which we will discuss in detail
in a separate publication. Our paper is organized as follows. In
section {\bf II}, we first discuss the thermal operator representation in
the imaginary time
formalism without a chemical potential and then with a
nontrivial chemical potential. In this section, we also point out
various properties of the thermal operator which are quite
interesting. In section {\bf III}, we derive the thermal operator
representation in the closed time path
formalism \cite{das:book97,Schwinger:1961qe,Bakshi:1962dvKeldysh:1964ud},
again with and without a
chemical potential, where the general proof is really much simpler
than the imaginary time formalism. In section {\bf IV}, we discuss 
the thermal operator representation for a general time contour, that includes
the case of thermofield dynamics \cite{das:book97,umezawa:1982nv},
where the thermal operator, in
general, is not as simple as in the imaginary time and the closed time
path formalisms. We conclude with a brief summary of our results in
section {\bf V}.

\section{Imaginary Time Formalism}

Let us consider a massive real scalar field theory in Euclidean
space. In this case, we know that the zero temperature propagator in
momentum space is given by
\begin{equation}
\Delta^{(T=0)} (p_E,E) = \frac{1}{p^{2} + m^{2}} = \frac{1}{p_{E}^{2} +
  E^{2}},\label{zeroTprop}
\end{equation}
where we have defined
\begin{equation}
E = \sqrt{\vec{p}^{\ 2} + m^{2}}.
\end{equation}
The energy variable can now be Fourier transformed to give the
propagator in the mixed space at zero temperature to be
\begin{equation}
\Delta^{(T=0)} (\tau, E) = \int \frac{\mathrm{d}p_{E}}{2\pi}\
\frac{e^{-ip_{E}\tau}}{p_{E}^{2} + E^{2}} = \frac{1}{2E}\left[\theta
  (\tau) e^{-E\tau} + \theta (-\tau) e^{E\tau}\right],\quad
-\infty\leq \tau\leq \infty.\label{zeroTpropt}
\end{equation}
At finite temperature in the imaginary time formalism (we will set the
Boltzmann constant to unity for simplicity), the propagator
in the momentum space has the same form as in (\ref{zeroTprop}) with
$p_{E} = 2\pi kT$ where $k$ is an integer. In this case, the Fourier
transform of the propagator leads to
\begin{eqnarray}
&  & \Delta^{(T)} (\tau, E) = T\sum_{k} \frac{e^{-i 2\pi k T  \tau}}
{(2\pi k T)^{2} + E^{2}}
 = \frac{1}{2E}\left[\left(\theta (\tau) + n (E)\right) e^{-E\tau}
   + \left(\theta(-\tau)+ n (E)\right) e^{E\tau}\right]\nonumber\\
 & &\quad = \frac{1}{2E}\left[\theta (\tau)\left\{(1+n(E))e^{-E\tau} + n(E)
     e^{E\tau}\right\} + \theta (-\tau)\left\{n(E) e^{-E\tau} +
     (1+n(E)) e^{E\tau}\right\}\right],\quad
 -\frac{1}{T}\leq \tau\leq \frac{1}{T},\label{Tprop}
\end{eqnarray}
which is symmetric under $\tau\rightarrow -\tau$. It is important to
recognize that, in the imaginary time formalism,
time is rotated to the negative imaginary axis and lies between the
interval $(0,\frac{1}{T})$ so that the propagator (being a difference
of time coordinates) is defined only within the interval
$(-\frac{1}{T}, \frac{1}{T})$. Furthermore, let us note that, in this
mixed space representation, the thermal propagator (\ref{Tprop}) can
be naturally written as the sum of a zero temperature part and a
finite temperature part much like in the real time formalisms
\cite{das:book97} 
\begin{equation}
\Delta^{(T)} (\tau, E) = \Delta^{(T=0)} (\tau,E) + \overline{\Delta}
(\tau,E),\label{Tdecomposition}
\end{equation}
where the finite temperature part has the form
\begin{equation}
\overline{\Delta} (\tau,E) = \frac{n(E)}{2E}\left(e^{-E\tau} +
  e^{E\tau}\right).
\end{equation}
Furthermore, we note that both the zero temperature propagator
(\ref{zeroTpropt}) and the finite temperature propagator (\ref{Tprop})
satisfy the same equation
\begin{equation}
\left(\frac{\partial^{2}}{\partial\tau^{2}} - E^{2}\right) \Delta
(\tau,E) = - \delta (\tau),\label{equation}
\end{equation}
but the finite temperature propagator satisfies the periodicity
condition 
\begin{equation}
\Delta^{(T)} (\tau<0,E) = \Delta^{(T)} (\tau +
\frac{1}{T},E),\label{kms}
\end{equation}
following from the KMS 
condition \cite{Kubo:1957xxMartin:1959jp}.
It follows, therefore, that
$\overline{\Delta}$ in (\ref{Tdecomposition}) satisfies the
homogeneous equation
\begin{equation}
\left(\frac{\partial^{2}}{\partial\tau^{2}} -
E^{2}\right)\overline{\Delta} (\tau,E) = 0,
\end{equation}
and is responsible for incorporating the periodicity condition
(\ref{kms}).

\begin{figure}[ht!]
\begin{center}
\includegraphics[scale=0.4]{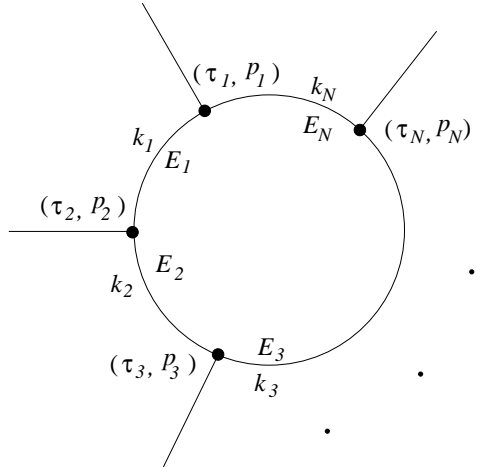}
\end{center}
\caption{One loop diagram in the $\phi^3$ theory with $N$ external
  time coordinates.} 
\label{fig1}
\end{figure}

It follows now from the forms of the propagators in (\ref{zeroTpropt})
and (\ref{Tprop}) that we can write
\begin{eqnarray}
\Delta^{(T)} (\tau,E) & = & \left(1 + n(E) (1 - S(E))\right)
\Delta^{(T=0)} (\tau, E) \nonumber \\
& = & {\cal O}^{(T)}(E)\,\Delta^{(T=0)}(\tau,E), \quad
-\frac{1}{T}\leq \tau\leq \frac{1}{T},\label{factorization}
\end{eqnarray}
where we note that the basic thermal operator ${\cal O}^{(T)} (E)$ is
independent of the time coordinate of the propagator. 
As we will see, this basic factorization of the finite temperature
propagator in terms of a thermal operator (that contains all the
temperature dependence, but no time) and the zero temperature
propagator (which carries all the time dependence) is at the heart of
the thermal operator representation of any finite temperature
graph. Given the factorization in (\ref{factorization}), it
immediately follows that any one loop graph with $N$ external lines
(see figure \ref{fig1})
would lead to the thermal operator representation (we consider the
$\phi^{3}$ theory for simplicity, neglect the
overall coupling constants, assume that all momenta are incoming and
identify $k_{N+1}=k_{1}, p_{N+1}=p_{1}$)

\begin{eqnarray}
& & \int \prod_{i=1}^{N} {\mathrm{d}^{3}k_{i}}\
\delta^{(3)}(k_i-k_{i+1}+p_{i+1})
\gamma_{N}^{(T)}\nonumber\\
 & &\quad = 
\int \prod_{i=1}^{N} {\mathrm{d}^{3}k_{i}}\
\delta^{(3)}(k_i-k_{i+1}+p_{i+1})  \Delta^{(T)}
(\tau_{1}-\tau_{2},E_{1})\cdots \Delta^{(T)}
(\tau_{N}-\tau_{1},E_{N})\nonumber\\
 & &\quad = \int \prod_{i=1}^{N} {\mathrm{d}^{3}k_{i}}\
\delta^{(3)}(k_i-k_{i+1}+p_{i+1})
\prod_{i=1}^{N} \left(1 + n_{i} (1 - S_{i})\right) \Delta^{(T=0)}
(\tau_{1}-\tau_{2},E_{1})\cdots \Delta^{(T=0)}
(\tau_{N}-\tau_{1},E_{N})\nonumber\\
& &\quad = \int\prod_{i=1}^{N} {\mathrm{d}^{3}k_{i}}\
\delta^{(3)}(k_i-k_{i+1}+p_{i+1})
{\cal O}^{(T)} \gamma_{N}^{(T=0)},
\end{eqnarray}
where we have identified
\begin{equation}
{\cal O}^{(T)} = \prod_{i=1}^{N} {\cal O}^{(T)} (E_{i}) =
\prod_{i=1}^{N} \left(1 + n_{i} (1-S_{i})\right).
\end{equation}

\begin{figure}[ht!]
\begin{center}
\includegraphics[scale=0.5]{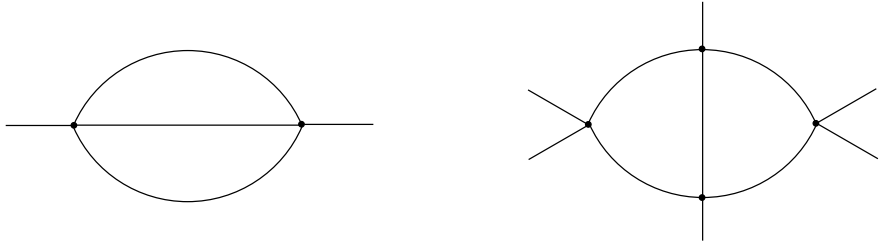}
\end{center}
\caption{Two loop diagrams in the $\phi^4$ theory without internal
  time coordinates.}
\label{fig2}  
\end{figure}

Similarly, the thermal operator representation immediately follows
from the factorization of the propagator (\ref{factorization}) for
any higher loop graph where all the vertices have only external times
(no internal vertices present). It is obvious, for example, in the
case of the graphs shown in figure \ref{fig2} in the $\phi^{4}$ theory. 

\begin{figure}[ht!]
\begin{center}
\includegraphics[scale=0.5]{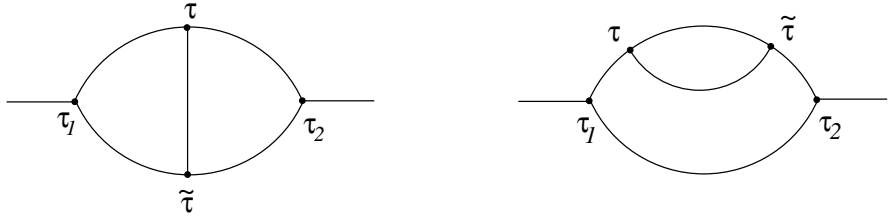}
\end{center}
\caption{Two loop self-energy diagrams in the $\phi^3$ theory with internal
  time coordinates $\tau,\tilde{\tau}$.}
\label{fig3}
\end{figure}

The difficulty in establishing the thermal operator representation for
an arbitrary graph arises when there are internal vertices present for
which the internal time coordinates have to be integrated over all
allowed values. For example, in the $\phi^{3}$ theory the self-energy
graph at two loops can have diagrams containing internal vertices of
the form shown in figure \ref{fig3}
($\tau,\tilde{\tau}$ in this case correspond to internal time
coordinates that have to be integrated over the allowed values).
Keeping in mind our earlier comments, we note that at
finite temperature, the time integration goes over
\begin{equation}
\int_{0}^{\frac{1}{T}} \mathrm{d}\tau,\label{Tintegration}
\end{equation}
while in zero temperature graphs, the internal time needs to be
integrated over
\begin{equation}
\int_{-\infty}^{\infty} \mathrm{d}\tau,\label{tintegration}
\end{equation}
and it is not clear {\em a priori} how the range of integration
(\ref{Tintegration}) can be extended to (\ref{tintegration}) in order
to establish the thermal operator relation. Furthermore, the finite
temperature propagator is only defined in the interval
$(-\frac{1}{T},\frac{1}{T})$, and it is not clear if it can be
analytically continued to other regions (preliminary analysis
indicates it cannot be done so consistently). Before giving a general
proof that such an extension of the range of integration can, in fact,
be made, let us work out
a simple example to bring out some of the essential features.

Let us first consider the product of $N$ propagators with a common
time that is being integrated, namely,
\begin{equation}
I_{1} = \int_{0}^{\frac{1}{T}} \mathrm{d}\tau\prod_{i=1}^{N}
\Delta^{(T)} (\tau-\tau_{i},E_{i}),
\end{equation}
where $\tau_{i}, i=1,2,\cdots , N$ are assumed to be external times
that lie between the interval $(0,\frac{1}{T})$. Using
(\ref{factorization}), we can write the above expression also as
\begin{eqnarray}
I_{1} & = & \prod_{i=1}^{N} \left(1 + n_{i} (1 - S_{i})\right)
\int_{0}^{\frac{1}{T}} \mathrm{d}\tau\prod_{i=1}^{N} \Delta^{(T=0)}
(\tau-\tau_{i},E_{i})\nonumber\\
 & = & \prod_{i=1}^{N}\left(1 + n_{i}
(1-S_{i})\right)\int_{-\infty}^{\infty} \mathrm{d}\tau
\prod_{i=1}^{N} \Delta^{(T=0)} (\tau-\tau_{i},E_{i}) -
\overline{I}_{1},
\end{eqnarray}
where we have defined
\begin{equation}
\overline{I}_{1} = \prod_{i=1}^{N} \left(1 + n_{i}
(1-S_{i})\right)\left[\int_{-\infty}^{0}\mathrm{d}\tau \prod_{i=1}^{N}
\Delta^{(T=0)} (\tau-\tau_{i},E_{i}) + \int_{\frac{1}{T}}^{\infty}
\mathrm{d}\tau\prod_{i=1}^{N} \Delta^{(T=0)}
(\tau-\tau_{i},E_{i})\right].\label{extra}
\end{equation}
We note here that it is because the basic thermal operator in the
factorization of the propagator is independent of the time coordinate
that it can be taken out of the time integral. Furthermore, 
 although the finite temperature propagator cannot be
extended beyond its domain, once we have extracted the thermal
operators, the zero temperature propagators are defined on the entire
real axis, a fact which we have used in the above. Since the external
times satisfy $0\leq \tau_{i}\leq \frac{1}{T}$, using the definition
of the zero temperature propagator in (\ref{zeroTpropt}) the integrals
in the bracket above can be evaluated in a simple manner and lead to
\begin{eqnarray}
& = & \left(\prod_{i=1}^{N} \frac{1}{2E_{i}}\right)
\left(\int_{-\infty}^{0}\mathrm{d}\tau\
e^{\sum_{i}E_{i}(\tau-\tau_{i})} +
\int_{\frac{1}{T}}^{\infty}\mathrm{d}\tau\ e^{-\sum_{i}E_{i}
(\tau-\tau_{i})}\right)\nonumber\\
&= & \left(\prod_{i=1}^{N}
\frac{1}{2E_{i}}\right)\frac{1}{\sum_{i}E_{i}} 
\left(e^{-\sum_{i}E_{i}\tau_{i}} 
+ e^{-\sum_{i} \frac{E_{i}}{T}}\times e^{\sum_{i}
E_{i}\tau_{i}}\right)\nonumber\\
& = & \frac{1}{\sum_{i}E_{i}}\left[\prod_{i=1}^{N}
\frac{e^{-E_{i}\tau_{i}}}{2E_{i}} +
\prod_{i=1}^{N} \frac{n_{i}}{1+n_{i}}
\frac{e^{E_{i}\tau_{i}}}{2E_{i}}\right]\nonumber\\
& = & \left(1 - \prod_{i=1}^{N} \frac{n_{i}}{1+n_{i}} (-S_{i})\right)
\frac{1}{\sum_{i}E_{i}} \prod_{i=1}^{N}
\frac{e^{-E_{i}\tau_{i}}}{2E_{i}}. \label{bracket}
\end{eqnarray}
It is now straightforward to show that the thermal operator
annihilates (\ref{bracket}), namely,
\begin{equation}
\prod_{i=1}^{N} \left(1 + n_{i}(1-S_{i})\right)\left(1 - \prod_{i=1}^{N}
 \frac{n_{i}}{1+n_{i}} (-S_{i})\right)\frac{1}{\sum_{i}E_{i}} \prod_{i=1}^{N}
\frac{e^{-E_{i}\tau_{i}}}{2E_{i}} =
0.\label{homogeneous}
\end{equation}
This, therefore, establishes that for a product of $N$ propagators
integrated over a single common time which is integrated, we can extend the range of
integration to write
\begin{equation}
\int_{0}^{\frac{1}{T}} \mathrm{d}\tau\prod_{i=1}^{N} \Delta^{(T)}
(\tau-\tau_{i},E_{i}) = \prod_{i=1}^{N}\left(1+n_{i} (1-S_{i})\right)
\int_{-\infty}^{\infty} \mathrm{d}\tau\prod_{i=1}^{N} \Delta^{(T=0)}
(\tau-\tau_{i},E_{i}). \label{singlet}
\end{equation}
(We want to emphasize here that even though the basic thermal operator
is independent of time, it is improper to take the products of the
thermal operators inside the integral in (\ref{singlet}) and write it
as a product of thermal operators being integrated over the interval
$(-\infty,\infty)$ since the thermal propagators are not defined
outside of their domain. Such an attempt would lead to various
divergences as well as inconsistency problems.)
The case when there is a single internal time integration works out in
a simple manner because the extra terms in (\ref{extra}) do not
involve any nontrivial time ordering as is clear from
(\ref{bracket}). This is no longer the case when there are two or more
internal time integrations. We can always extend one of the time
integrations to the entire real axis as discussed above. But, once
this is done, the subsequent integrals will involve nontrivial time
ordering. Nonetheless, case by case, one finds explicitly that the integration
range can be extended to the entire real axis (after factoring out
the thermal operator) when two internal time integrations are
involved. This, in turn, suggests that there must be a general proof
for such an equivalence for an arbitrary number of internal time
integrations which we discuss next.

\subsection{General Proof}

We have already seen that when there is one internal time coordinate
that is being integrated, the range of the integration can be extended
from $(0,\frac{1}{T})$ to $(-\infty,\infty)$ under the action of the
thermal operator. The generalization of this result to an arbitrary
number of internal times that are being integrated can be carried out
in the following way. First, let us note that since the zero
temperature (as well as the finite temperature) propagator satisfies
(\ref{equation}), it follows that
\begin{eqnarray}
& & \left(-\frac{\partial^{2}}{\partial\tau_{k}^{2}} + E_{k}^{2}\right)
\int_{-\infty}^{\infty} \mathrm{d}\tau \prod_{i=1}^{N} \Delta^{(T=0)}
(\tau-\tau_{i},E_{i})\nonumber\\
 & & \quad = \int_{-\infty}^{\infty} \mathrm{d}\tau
 \sideset{}{^\prime}\prod_{i=1}^{N}  
\Delta^{(T=0)}
(\tau-\tau_{i},E_{i})\left(-\frac{\partial^{2}}{\partial\tau_{k}^{2}}
+ E_{k}^{2}\right)\Delta^{(T=0)} (\tau-\tau_{k},E_{k})\nonumber\\
& & \quad = \int_{-\infty}^{\infty} \mathrm{d}\tau
\sideset{}{^\prime}\prod_{i=1}^{N}  
\Delta^{(T=0)} (\tau-\tau_{i},E_{i}) \delta (\tau-\tau_{k})\nonumber\\
& & \quad = \sideset{}{^\prime}\prod_{i=1}^{N}   \Delta^{(T=0)}
(\tau_{k}-\tau_{i},E_{i}).
\end{eqnarray}
Here $\tau_{k}$ is any external time coordinate and the prime on the
product implies the absence of the term with $i=k$. As a result of
this identity, we can write
\begin{equation}
\int_{-\infty}^{\infty} \mathrm{d}\tau\prod_{i=1}^{N} \Delta^{(T=0)}
(\tau-\tau_{i},E_{i}) =
\left(-\frac{\partial^{2}}{\partial\tau_{k}^{2}} +
E_{k}^{2}\right)^{-1} \left(\sideset{}{^\prime}\prod_{i=1}^{N}   \Delta^{(T=0)}
(\tau_{k}-\tau_{i},E_{i})\right).\label{integration}
\end{equation}

In general, a homogeneous term (a term annihilated by the differential
operator) is allowed on the right hand side of
(\ref{integration}). However, since a Feynman diagram is a time
ordered quantity, a homogeneous term is not expected on physical
grounds. That this is true mathematically can also be seen as
follows. Let us evaluate the integral (\ref{integration}) in the
momentum space. Using the
definition of the Euclidean propagator in momentum space in
(\ref{zeroTprop}), we obtain
\begin{eqnarray}
& & \int_{-\infty}^{\infty} \mathrm{d}\tau\prod_{i=1}^{N}
\Delta^{(T=0)} (\tau-\tau_{i},E_{i})\nonumber\\
& & \quad = \int
\mathrm{d}\tau\prod_{i=1}^{N}\frac{\mathrm{d}p_{iE}}{2\pi}
\prod_{i=1}^{N}\frac{e^{-ip_{iE}(\tau-\tau_{i})}}{(p_{iE}^{2} +
E_{i}^{2})}\nonumber\\
& & \quad = \int \prod_{i=1}^{N}\frac{\mathrm{d}p_{iE}}{2\pi}\
2\pi\delta (p_{1E}+\cdots + p_{NE}) \prod_{i=1}^{N}
\frac{e^{ip_{iE}\tau_{i}}}{(p_{iE}^{2} + E_{i}^{2})}.
\end{eqnarray}
The delta function allows us to eliminate one of the $p_{iE}$
variables and let us choose the dependent variable to be
$p_{kE}$ related to $\tau_{k}$. Eliminating this variable, we obtain
\begin{eqnarray}
& & \int_{-\infty}^{\infty} \mathrm{d}\tau\prod_{i=1}^{N}
\Delta^{(T=0)} (\tau-\tau_{i},E_{i})\nonumber\\
& & \quad = \int \sideset{}{^\prime}\prod_{i=1}^{N}
\frac{\mathrm{d}p_{iE}}{2\pi}\ 
\frac{1}{\left((\sum_{i}^{\prime}  p_{iE})^{2} + E_{k}^{2}\right)}
\sideset{}{^\prime}\prod_{i=1}^{N}
\frac{e^{-ip_{iE}(\tau_{k}-\tau_{i})}}{(p_{iE}^{2} +
E_{i}^{2})} \nonumber\\
& & \quad = \left(-\frac{\partial^{2}}{\partial\tau_{k}^{2}} +
E_{k}^{2}\right)^{-1} \int \sideset{}{^\prime}\prod_{i=1}^{N}  
\frac{\mathrm{d}p_{iE}}{2\pi} \sideset{}{^\prime}\prod_{i=1}^{N}   \frac{e^{-ip_{iE}
(\tau_{k}-\tau_{i})}}{(p_{iE}^{2} + E_{i}^{2})}\nonumber\\
& & \quad = \left(-\frac{\partial^{2}}{\partial\tau_{k}^{2}} +
E_{k}^{2}\right)^{-1} \sideset{}{^\prime}\prod_{i=1}^{N}   \Delta^{(T=0)}
(\tau_{k}-\tau_{i},E_{i}),
\end{eqnarray}
which is the result obtained in (\ref{integration}). We would like to
note here that the operator
$\left(-\frac{\partial^{2}}{\partial\tau_{k}^{2}} +
E_{k}^{2}\right)^{-1}$ can be thought of in terms of the standard integral
representation as
\begin{eqnarray}
\left(-\frac{\partial^{2}}{\partial\tau_{k}^{2}} +
E_{k}^{2}\right)^{-1} & = & \int_{0}^{\infty} \mathrm{d}\alpha\ e^{-\alpha
(-\frac{\partial^{2}}{\partial\tau_{k}^{2}} + E_{k}^{2})}\nonumber\\
 & = & \sum_{N=0}^{\infty} \int_{0}^{\infty} \mathrm{d}\alpha\ \frac{e^{-\alpha
E_{k}^{2}}}{N!}
\alpha^{N}\frac{\partial^{2N}}{\partial\tau_{k}^{2N}}\nonumber\\
 & = & \frac{1}{E_{k}^{2}}\left[1 + \sum_{N=1}^{\infty} \left(\frac{1}{E_{k}^{2}}
\frac{\partial^{2}}{\partial\tau_{k}^{2}}\right)^{N}\right].
\end{eqnarray}

The relation (\ref{integration}) is quite useful in proving that the
range of the finite temperature integration can be extended to the
entire real axis under the action of the thermal operator for any
number of internal time integrations. The important thing to note here
is that the differential operator commutes with the thermal operator
(since it is independent of time)
and that the equivalence can be established recursively as
follows. First, we note from (\ref{singlet}) that for a single time
integration this is true. Let us assume that we have a product of
propagators with two internal times that are integrated. The most
general form for such a product can be written as
\begin{eqnarray}
& & \int_{0}^{\frac{1}{T}} \mathrm{d}\tau \mathrm{d}\tilde{\tau}
\left(\prod_{i=1}^{N_{1}} \Delta^{(T)} (\tau-\tau_{i},E_{i})\right)
\left(\prod_{\alpha=1}^{N_{2}} \Delta^{(T)}
(\tau-\tilde{\tau},\tilde{E}_{\alpha})\right)\left(\prod_{\mu=1}^{N_{3}}
\Delta^{(T)}
(\tilde{\tau}-\bar{\tau}_{\mu},\bar{E}_{\mu})\right)\nonumber\\
& & \quad = {\cal O}^{(T)} \int_{0}^{\frac{1}{T}} \mathrm{d}\tau
\mathrm{d}\tilde{\tau}
\left(\prod_{i=1}^{N_{1}} \Delta^{(T=0)} (\tau-\tau_{i},E_{i})\right)
\left(\prod_{\alpha=1}^{N_{2}} \Delta^{(T=0)}
(\tau-\tilde{\tau},\tilde{E}_{\alpha})\right)\left(\prod_{\mu=1}^{N_{3}}
\Delta^{(T=0)}
(\tilde{\tau}-\bar{\tau}_{\mu},\bar{E}_{\mu})\right)\nonumber\\
& & \quad = {\cal O}^{(T)}
\int_{-\infty}^{\infty}\mathrm{d}\tau\int_{0}^{\frac{1}{T}}
\mathrm{d}\tilde{\tau}\left(\prod_{i=1}^{N_{1}} \Delta^{(T=0)}
(\tau-\tau_{i},E_{i})\right)
\left(\prod_{\alpha=1}^{N_{2}} \Delta^{(T=0)}
(\tau-\tilde{\tau},\tilde{E}_{\alpha})\right)\left(\prod_{\mu=1}^{N_{3}}
\Delta^{(T=0)}
(\tilde{\tau}-\bar{\tau}_{\mu},\bar{E}_{\mu})\right)\nonumber\\
& & \quad = {\cal O}^{(T)}  \left(-\frac{\partial^{2}}{\partial\tau_{1}^{2}}
+ E_{1}^{2}\right)^{-1}\!\!\! 
\left(\prod_{i=2}^{N_{1}} \Delta^{(T=0)} (\tau_{1}-\tau_{i},E_{i})\!\right)\!\!
\int_{0}^{\frac{1}{T}} \mathrm{d}\tilde{\tau} 
\left(\prod_{\alpha=1}^{N_{2}} \Delta^{(T=0)}
(\tau_{1}-\tilde{\tau},\tilde{E}_{\alpha})\!\right)\!\!\left(\prod_{\mu=1}^{N_{3}}
\Delta^{(T=0)}
(\tilde{\tau}-\bar{\tau}_{\alpha},\bar{E}_{\mu})\!\right),\nonumber\\
& & 
\end{eqnarray}
where $N_{1},N_{2},N_{3}$ are arbitrary integers.
Here we have identified
\begin{equation}
{\cal O}^{(T)} = \left(\prod_{i=1}^{N_{1}}
\left(1+n_{i}(1-S_{i})\right)\right)\left(\prod_{\alpha=1}^{N_{2}}\left(1
+ \tilde{n}_{\alpha}
(1-\tilde{S}_{\alpha})\right)\right)\left(\prod_{\mu=1}^{N_{3}}
\left(1+\bar{n}_{\mu} (1-\bar{S}_{\mu})\right)\right),
\end{equation}
as well as used (\ref{integration}) with $\tau_{1}$ as the external
time coordinate for simplicity. This shows that the case of two
internal time integrations can be reduced to that of a single time
integration where we know from (\ref{singlet}) that the range of
integration can be extended to the entire real axis (under the action
of the thermal operator) so that we have
\begin{eqnarray}
& & \int_{0}^{\frac{1}{T}} \mathrm{d}\tau \mathrm{d}\tilde{\tau}
\left(\prod_{i=1}^{N_{1}} \Delta^{(T)} (\tau-\tau_{i},E_{i})\right)
\left(\prod_{\alpha=1}^{N_{2}} \Delta^{(T)}
(\tau-\tilde{\tau},\tilde{E}_{\alpha})\right)\left(\prod_{\mu=1}^{N_{3}}
\Delta^{(T)}
(\tilde{\tau}-\bar{\tau}_{\alpha},\bar{E}_{\mu})\right)\nonumber\\
& & \quad = {\cal O}^{(T)}  \left(-\frac{\partial^{2}}{\partial\tau_{1}^{2}}
+ E_{1}^{2}\right)^{-1}\!\! 
\left(\prod_{i=2}^{N_{1}} \Delta^{(T=0)} (\tau_{1}-\tau_{i},E_{i})\right)\!\!
\int_{-\infty}^{\infty}\! \mathrm{d}\tilde{\tau} 
\left(\prod_{\alpha=1}^{N_{2}} \Delta^{(T=0)}
(\tau_{1}-\tilde{\tau},\tilde{E}_{\alpha})\right)\!\!\left(\prod_{\mu=1}^{N_{3}}
\Delta^{(T=0)}
(\tilde{\tau}-\bar{\tau}_{\alpha},\bar{E}_{\mu})\right)\nonumber\\
& & \quad = {\cal O}^{(T=0)} \int_{-\infty}^{\infty} \mathrm{d}\tau \mathrm{d}\tilde{\tau}
\left(\prod_{i=1}^{N_{1}} \Delta^{(T=0)} (\tau-\tau_{i},E_{i})\right)
\left(\prod_{\alpha=1}^{N_{2}} \Delta^{(T=0)}
(\tau-\tilde{\tau},\tilde{E}_{\alpha})\right)\left(\prod_{\mu=1}^{N_{3}}
\Delta^{(T=0)}
(\tilde{\tau}-\bar{\tau}_{\alpha},\bar{E}_{\mu})\right),
\end{eqnarray}
where we have used (\ref{singlet}) to restore the $\tau$
integration. This process can be used recursively to show that the
range of integration can be extended to the entire real axis (under
the action of the thermal operator) for any number of internal time
integrations. This, therefore, proves the thermal operator
representation for any arbitrary graph with $N$ external legs, namely,
\begin{equation}
\int \prod_{i=1}^{I} \frac{\mathrm{d}^{3}k_{i}}{(2\pi)^{3}}
\prod_{v=1}^{V} {(2\pi)^{3}}
\delta_{v}^{(3)}(k,p)
\gamma_{N}^{(T)} = \int \prod_{i=1}^{I}
\frac{\mathrm{d}^{3}k_{i}}{(2\pi)^{3}} 
\prod_{v=1}^{V} {(2\pi)^{3}} \delta_{v}^{(3)}(k,p)
{\cal O}^{(T)}  \gamma_{N}^{(T=0)}.\label{tor}
\end{equation}

\begin{figure}[ht!]
\begin{center}
\includegraphics[scale=0.5]{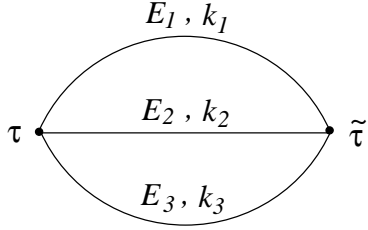}
\end{center}
\caption{Pressure diagram in the $\phi^3$ theory at two loops.}
\label{fig4}  
\end{figure}

There are several things to note here. The proof of the thermal
operator representation in the mixed space is more direct and the
origin of this relation can be traced to the factorization of the
thermal propagator in terms of the basic thermal operator which is
independent of time and the zero
temperature propagator. There is no necessity for classifying the
graphs into {\em trees} or for introducing any regularization as one
does in the momentum space analysis (which is quite unusual since the
finite temperature results are not expected to be divergent). As we
will show later, the derivation of the thermal operator representation
in the mixed space is even simpler in the closed time path
formalism.  We also note here that although this analysis seems to
suggest that this equivalence holds only for graphs with external
legs, such a relation holds even for graphs without any external legs
which follows in a straight forward manner from the closed time
path formalism to be discussed later. Here we simply note that if we
are looking at the pressure in the $\phi^{3}$ theory at two loops at
finite temperature (see figure \ref{fig4})
we have the explicit result (neglecting the overall factors
involving the coupling as well as the symmetry factor)
\begin{equation}
P^{(T)}  = \int \left(\prod_{i=1}^{3} \frac{\mathrm{d}^{3}k_{i}}{(2\pi)^{3}}\right)
{(2\pi)^{3}} \delta^{(3)}(k_{1}+k_{2}+k_{3}) {(2\pi)^{3}} \delta^{(3)}(0) J^{(T)},
\end{equation}
where
\begin{eqnarray}
J^{(T)} & = & \int_{0}^{\frac{1}{T}} \mathrm{d}\tau
  \mathrm{d}\tilde{\tau} \Delta^{(T)} (\tau-\tilde{\tau},E_{1})
  \Delta^{(T)} (\tau-\tilde{\tau},E_{2})\Delta^{(T)}
  (\tau-\tilde{\tau},E_{3})\nonumber\\
 & = & {\cal O}^{(T)}  \int_{0}^{\frac{1}{T}}
 \mathrm{d}\tau\mathrm{d}\tilde{\tau}\Delta^{(T=0)} (\tau-\tilde{\tau},E_{1})
  \Delta^{(T=0)} (\tau-\tilde{\tau},E_{2})\Delta^{(T=0)}
  (\tau-\tilde{\tau},E_{3}) \nonumber\\
 & = & {\cal O}^{(T)}  \int_{0}^{\frac{1}{T}} \mathrm{d}\tau
 \mathrm{d}\tilde{\tau}\
 \frac{2(E_{1}+E_{2}+E_{3})}{(2E_{1})(2E_{2})(2E_{3})}  \Delta^{(T=0)}
 (\tau-\tilde{\tau},E_{1}+E_{2}+E_{3}).
\end{eqnarray}
The thermal operator for this graph is given by
\begin{equation}
{\cal O}^{(T)}  = \prod_{i=1}^{3} \left(1 + n_{i} (1-S_{i})\right).
\end{equation}
As we have discussed earlier, in the case of a single integration over
time, the range of integration can be extended to the entire real axis
under the action of the thermal operator. Using this as well as
shifting variables of integration, we obtain
\begin{eqnarray}
J^{(T)} & = & {\cal O}^{(T)}  \int_{-\infty}^{\infty} \mathrm{d}\tau
\int_{0}^{\frac{1}{T}} \mathrm{d}\tilde{\tau}\ 
\frac{2(E_{1}+E_{2}+E_{3})}{(2E_{1})(2E_{2})(2E_{3})} \Delta^{(T=0)}
(\tau-\tilde{\tau},E_{1}+E_{2}+E_{3}) \nonumber\\
 & = & {\cal O}^{(T)}  \int_{0}^{\frac{1}{T}}\mathrm{d}\tilde{\tau}
 \int_{-\infty}^{\infty} \mathrm{d}\tau\
 \frac{2(E_{1}+E_{2}+E_{3})}{(2E_{1})(2E_{2})(2E_{3})} \Delta^{(T=0)}
 (\tau,E_{1}+E_{2}+E_{3})\nonumber\\
 & = & {\cal O}^{(T)}  \frac{1}{T} \delta_{k_{0}=0,0} \int_{-\infty}^{\infty}
 \mathrm{d}\tau\ \frac{2(E_{1}+E_{2}+E_{3})}{(2E_{1})(2E_{2})(2E_{3})}
 \Delta^{(T=0)} (\tau,E_{1}+E_{2}+E_{3}),
\end{eqnarray}
where we have used the basic definition of Fourier transform for a
finite interval ($k_{0}$ is an integer)
\begin{equation}
\int_{0}^{\frac{1}{T}} \mathrm{d}\tilde{\tau}\ e^{-ik_{0}\tilde{\tau}}
= \frac{1}{T}\ \delta_{k_{0},0}.
\end{equation}
Recalling that in the continuum limit
\begin{equation}
\frac{1}{T} \delta_{k_{0}=0,0}\rightarrow 2\pi
\delta(0),\label{identification}
\end{equation}
we immediately identify that
\begin{equation}
J^{(T)} = {\cal O}^{(T)}  J^{(T=0)},
\end{equation}
and the thermal operator representation works even for graphs without
any external legs. This is more directly seen in the closed time path
formalism that we will discuss later. It is worth noting here that in
the imaginary time formalism, the graphs without any external leg
always have a factor $\frac{1}{T} \delta_{k_{0}=0,0}$ which must be
identified with the continuum case as in (\ref{identification}). This
is not necessary in the real time formalism where time is a continuous
variable defined over the entire real axis.

Let us also comment here on some interesting aspects of calculations
in the mixed space. The calculations in this case are more like the
real time calculations in the sense that the amplitudes contain all
possible factors of the statistical distribution function. However,
when the 
results are Fourier transformed into energy-momentum space, because of
various identities, the number of statistical distribution functions in an
amplitude reduce to one per internal loop
\cite{Espinosa:2003af,Espinosa:2005gq}. We note that given the
thermal operator representation (\ref{tor}) of a graph in the mixed space, one
can go to the energy-momentum space in the standard manner. Here, we
have to simply remember that the zero temperature amplitude
$\gamma_{N}^{(T=0)} (E_{i},\tau_{\alpha})$ is a function of external
times which are restricted to lie between $0\leq \tau_{\alpha}\leq
\frac{1}{T}$. As a result, the Fourier transform needed to go to the
energy-momentum space is that over a finite interval (involving integer
energy) even though we have a zero temperature amplitude in the
Euclidean space.

The thermal operator representation for any graph at finite
temperature is a remarkable result. Physically, a Feynman graph at
finite temperature represents an ensemble average while a zero
temperature graph corresponds to a vacuum expectation
value. Therefore, a relation between the two can exist only if the
expectation value for a string of operators in any complete set of
states  (say in the energy
eigenbasis) would be proportional to the vacuum expectation value of
the same string of operators. Although at first sight
this seems unlikely, let us show
that this is plausible with the simple example of the propagator for a
massive, real scalar field at the
tree level (which will also explain the factorization for the
thermal propagator). Using the standard field decomposition in the Euclidean
space, we note that we can write (in the mixed space)
\begin{equation}
\phi (\tau,\vec{p}\ ) = \frac{1}{\sqrt{2E}}\left(e^{-E\tau}\
  a(\vec{p}\ ) + e^{E\tau}\ a^{\dagger} (-\vec{p}\ )\right),
\end{equation}
where as before, $E=\sqrt{\vec{p}^{\ 2}+ m^{2}}$. At zero temperature,
this leads to
\begin{equation}
\langle
0|T\left(\phi(\tau_{1},\vec{p}_{1})\phi(\tau_{2},\vec{p}_{2})\right)|
0\rangle = \delta^{(3)} (\vec{p}_{1}+\vec{p}_{2})\
\frac{1}{2E_{1}}\left(\theta(\tau_{1}-\tau_{2})
  e^{-E_{1}(\tau_{1}-\tau_{2})} + \theta(\tau_{2}-\tau_{1})
  e^{E_{1}(\tau_{1}-\tau_{2})}\right).
\end{equation}
On the other hand, in any eigenstate of energy containing $N$ quanta
of momentum $\vec{p}$, we have the expectation value
\begin{eqnarray}
&  & \langle N,\vec{p}\ | T\left(\phi(\tau_{1},\vec{p}_{1}\
  )\phi(\tau_{2},\vec{p}_{2}\ )\right)|N,\vec{p}\ \rangle\nonumber\\
& &\quad = 
\frac{1}{\sqrt{4E_{1}E_{2}}} \langle
N,\vec{p}\ |\left[\theta(\tau_{1}-\tau_{2})\left(
a(\vec{p}_{1})a^{\dagger}(-\vec{p}_{2})
e^{-E_{1}\tau_{1}+E_{2}\tau_{2}} + a^{\dagger}
(-\vec{p}_{1})a(\vec{p}_{2})
e^{E_{1}\tau_{1}-E_{2}\tau_{2}}\right)\right.\nonumber\\
 &  & \qquad \left. +
   \theta(\tau_{2}-\tau_{1})\left(a(\vec{p}_{2})a^{\dagger}
     (-\vec{p}_{1}) e^{E_{1}\tau_{1}-E_{2}\tau_{2}} + a^{\dagger}
     (-\vec{p}_{2})a(\vec{p}_{1})
     e^{-E_{1}\tau_{1}+E_{2}\tau_{2}}\right)\right]|N,\vec{p}\ \rangle\nonumber\\
 & &\quad = \delta^{(3)} (\vec{p}_{1}+\vec{p}_{2}) \delta^{(3)}
 (\vec{p}-\vec{p}_{1}) \frac{1}{2E_{1}}\left[\theta(\tau_{1}-\tau_{2})
   \left((1+N) e^{-E_{1} (\tau_{1}-\tau_{2})} + N
     e^{E_{1}(\tau_{1}-\tau_{2})}\right)\right.\nonumber\\
 &  & \qquad \left. + \theta (\tau_{2}-\tau_{1}) \left(N
     e^{-E_{1}(\tau_{1}-\tau_{2})} + (1 +N)
     e^{E_{1}(\tau_{1}-\tau_{2})}\right)\right]\nonumber\\
 &  &\quad = \delta^{(3)} (\vec{p}_{1}+\vec{p}_{2}) \delta^{(3)}
 (\vec{p}-\vec{p}_{1}) \left(1 + N (1 - S(E_{1})\right)
 \frac{1}{2E_{1}}\left[\theta(\tau_{1}-\tau_{2})
   e^{-E_{1}(\tau_{1}-\tau_{2})} + \theta (\tau_{2}-\tau_{1}) e^{E_{1}
     (\tau_{1}-\tau_{2})}\right]\nonumber\\
&  &\quad = \delta^{(3)} (\vec{p}-\vec{p}_{1}) \left(1 + N (
  1-S(E_{1})\right) \langle
0|T\left(\phi(\tau_{1},\vec{p}_{1})\phi(\tau_{2},\vec{p}_{2})\right)|0\rangle
.
\end{eqnarray}
This shows that the expectation value of the time ordered product of
two fields in any higher energy state is proportional to the
expectation value of the same operators in the vacuum state and
the proportionality factor is reminiscent of the basic thermal operator. In
fact, carrying out the thermal ensemble average, this proportionality
factor indeed becomes the basic thermal operator of
(\ref{factorization}). 

\subsection{Some Properties of the Thermal Operator}

The basic thermal operator that leads to factorization of the thermal
propagator has several interesting features. In this section, we
discuss some of them that are relevant to a better understanding of
this 
factorization. This will also be quite useful in connection with the
study of factorization in the real time formalisms. Let us note some
of the basic properties of the reflection operator. By definition
$(S(E))^{2} = 1$ and for the bosonic distribution functions we have
(this is identical to the well known result $n(-E) = - (1+n(E))$ for a
bosonic distribution function) 
\begin{eqnarray}
S(E) n (E) & = & S(E) \frac{1}{e^{\frac{E}{T}} -1} = \frac{1}{e^{-\frac{E}{T}} - 1}
S(E) = - (1 + n(E)) S(E),\nonumber\\
S(E) (1 + n(E)) & = & S(E) \frac{e^{\frac{E}{T}}}{e^{\frac{E}{T}} -1} =
\frac{e^{-\frac{E}{T}}}{e^{-\frac{E}{T}}-1} S(E) = - n(E)S(E).
\end{eqnarray}
Using these basic properties, it is easy to see that the basic thermal
operator can be written in various ways as
\begin{equation}
{\cal O}^{(T)}  (E) = 1 + n(E) (1-S(E)) = (1+S(E))(1+n(E)) =
-(1+S(E))n(E)S(E).\label{definitions}
\end{equation}
Furthermore, it follows from the definition that
\begin{eqnarray}
\left({{\cal O}^{(T)}}(E)\right)^2 & = & (1+ n - nS)(1+n-nS) \nonumber\\
& = & (1+n)^{2} - n(1+n) + \left(n^{2} - n(1+n)\right)S\nonumber\\
& = & \left(1+ n - nS\right) = {\cal O}^{(T)}  (E).\label{projection1}
\end{eqnarray}
The basic thermal operator, therefore, is a projection
operator. Consequently, the inverse of this operator does not exist
and the thermal operator representation for graphs does not have an
inverse relation. We note from (\ref{definitions}) that
\begin{equation}
S(E) {\cal O}^{(T)}(E) = {\cal O}^{(T)} (E),
\end{equation}
so that we have
\begin{equation}
\frac{1-S(E)}{2} {\cal O}^{(T)}  (E) = 0,\quad \frac{1+S(E)}{2} 
{\cal O}^{(T)} (E) = {\cal O}^{(T)}(E).
\end{equation}

Let us note that we can define another operator
\begin{equation}
{\overline{\cal O}}^{(T)}(E) = 1 + n(E) (1+ S(E)),
\end{equation}
which also corresponds to a projection operator and satisfies
\begin{equation}
\left({\overline{\cal O}}^{(T)}(E)\right)^2 = \overline{\cal O}^{(T)}(E),\quad
\frac{(1+S(E))}{2}\overline{\cal O}^{(T)}(E) = 0,\quad
\frac{(1-S(E))}{2}\overline{\cal O}^{(T)}(E) =  \overline{\cal O}^{(T)}(E).
\end{equation}
The two projection operators, however, are not orthogonal and satisfy
\begin{equation}
\overline{{\cal O}}^{(T)} (E) {{\cal O}}^{(T)}(E) = (1+2n(E)) {{\cal
    O}}^{(T)}(E),\quad {\cal O}^{(T)}(E)\overline{{\cal O}}^{(T)}(E) =
(1+2n(E)) \overline{{\cal O}}^{(T)}(E).
\end{equation}
The thermal operators, of course, depend on the temperature. Denoting
the temperature dependence explicitly, we can write
\begin{equation}
{{\cal O}}^{(T)} (E) = 1+ n(E,T) (1-S(E)).
\end{equation}
It can now be directly checked that
\begin{equation}
{{\cal O}}^{(T_2)}(E) {{\cal O}}^{(T_1)}(E) = {{\cal O}}^{(T_1)}(E).
\end{equation}
At first sight this may seem a bit strange. However, this result is
quite consistent with the underlying physics. Let us recall that the
effect of the thermal operator is to reproduce an ensemble
average. Once the ensemble average has been done through the operator
${{\cal O}}^{(T_1)}(E)$, the resulting amplitude is a scalar (proportional
to the identity operator). The application of a second thermal
operator ${{\cal O}}^{(T_2)}(E)$ to the result is then equivalent to
thermal averaging the identity operator which simply results in a
multiplicative factor of unity.

Finally, to understand the meaning of the thermal operator 
${{\cal O}}^{(T)}(E)$ as a projection operator, let us note from
(\ref{Tdecomposition}) as well as (\ref{factorization}) that
\begin{equation}
{{\cal O}}^{(T)}(E) \Delta^{(T=0)} (\tau,E) = \Delta^{(T)} (\tau,E) =
\Delta^{(T=0)} (\tau,E) + \overline{\Delta}(\tau,E). 
\end{equation}
We know that the zero temperature propagator does not satisfy the
periodicity condition (\ref{kms}). Rather, it is the temperature
dependent term $\overline{\Delta} (\tau,E)$ that enforces the
periodicity condition. Thus, we can think of the thermal operator as
projecting on to the space of functions satisfying the periodicity
condition. Of course, it follows from the definition that
\begin{equation}
{{\cal O}}^{(T)}(E) \overline{\Delta}(\tau,E) = 0,
\end{equation}
which is consistent with the fact that $\overline{\Delta}(\tau,E)$ is
the homogeneous solution of the Green's function equation
(\ref{equation}) (recall that the thermal operator commutes with the
differential operator). The meaning of the projection operator
$\overline{\cal O}^{(T)} (E)$ is also now clear since it can be
directly checked that
\begin{equation}
\overline{\cal O}^{(T)} (E) \overline{\Delta} (\tau, E) = 2 (1+n (E))
\overline{\Delta} (\tau,E).
\end{equation}

\subsection{Chemical Potential}

So far our discussion has been within the context of a canonical
ensemble where there is no chemical potential. Chemical potentials
arise when we have a conserved charge and we are dealing with a grand
canonical ensemble. In this case, the Hamiltonian in the
definition of the partition function is 
generalized to \cite{lebellac:book96}
\begin{equation}
 H\rightarrow  H - \mu Q,\label{grandcanonical}
\end{equation}
where $Q$ represents the conserved charge and $\mu$ the chemical
potential associated with it. In the context of a scalar field we can
introduce a chemical potential if we are dealing with a complex scalar
field where there is a natural definition of a conserved charge. The
free Lagrangian density, for such a theory in the presence of a
chemical potential, can be written as
\begin{equation}
{\cal L} = \left(\partial_{t} + i\mu\right)\phi^{*}
\left(\partial_{t}-i\mu\right)\phi  - \vec{\nabla}\phi^{*}\cdot
\vec{\nabla}\phi - m^{2} \phi^{*}\phi.\label{lagrangian}
\end{equation}
It can be checked that the Lagrangian density (\ref{lagrangian}) leads
to the Hamiltonian (\ref{grandcanonical}) where $Q$ represents the
conserved charge that generates the global phase transformations of
the system. We note that the addition of a chemical potential can be
viewed as 
introducing a constant electrostatic potential into the system. We
also recall here that for a relativistic massive boson, the chemical
potential has to satisfy
\begin{equation}
\mu \leq m.\label{condition}
\end{equation}

In Euclidean space, the Lagrangian density (\ref{lagrangian}) takes
the form
\begin{equation}
{\cal L}_{E} = \left(\partial_{\tau}+\mu\right)\phi^{*}
\left(\partial_{\tau}-\mu\right)\phi + \vec{\nabla}\phi^{*}\cdot
\vec{\nabla}\phi + m^{2}\phi^{*}\phi,
\end{equation}
which leads to the zero temperature propagator in momentum space to be
\begin{equation}
\Delta^{(T=0,\mu)} (p_{E},E) = \frac{1}{(p_{E} - i\mu)^{2} +
E^{2}},\label{zeroTmuprop}
\end{equation}
where as before we have
\begin{equation}
E^{2} = \vec{p}^{\;2} + m^{2}.
\end{equation}
This can be Fourier transformed to give the mixed space representation
of the zero temperature propagator to be
\begin{eqnarray}
\Delta^{(T=0,\mu)} (\tau,E) & = & \int \frac{\mathrm{d}p_{E}}{2\pi}\
\frac{e^{-ip_{E}\tau}}{(p_{E}-i\mu)^{2} + E^{2}}\nonumber\\
 & = & \frac{1}{2E}\left[\theta(\tau) e^{-(E-\mu)\tau} + \theta (-\tau)
e^{(E+\mu)\tau}\right].\label{zeroTmupropt}
\end{eqnarray}
At finite temperature, the momentum space propagator continues to be
given by (\ref{zeroTmuprop}) with $p_{E} = 2\pi kT$ where $k$ is an
integer. Fourier transforming this, we obtain the mixed space
representation for the thermal propagator to be
\begin{eqnarray}
\Delta^{(T,\mu)} (\tau,E) & = & T \sum_{k} \frac{e^{-2i\pi
kT \tau}}{(2\pi kT - i\mu)^{2} + E^{2}}\nonumber\\
 & = & \frac{1}{2E}\left[(\theta (\tau) + n (E-\mu)) e^{-(E-\mu)\tau} +
(\theta (-\tau) + n (E+\mu)) e^{(E+\mu)\tau}\right]\nonumber\\
 & = & \frac{1}{2E}\left[\theta(\tau)\left\{(1 + n_{-})e^{-(E-\mu)\tau}
+ n_{+} e^{(E+\mu)\tau}\right\} + \theta (-\tau)\left\{n_{-}
e^{-(E-\mu)\tau} + (1 +
n_{+})e^{(E+\mu)\tau}\right\}\right],\label{Tmuprop}
\end{eqnarray}
where, for simplicity of notation we have defined
\begin{equation}
n_{\pm} = n (E\pm \mu).\label{distribution}
\end{equation}
We note here that when $\mu=0$, this reduces to (\ref{Tprop}) as it
should. However, unlike the real scalar field, here the
propagator carries a direction, namely, the direction of the charge
flow (from $\phi$ to $\phi^{*}$). The propagator is not symmetric
under $\tau\leftrightarrow -\tau$. This is a reflection of the fact
that the chemical potential inherently distinguishes between particles
and anti-particles. However, under the simultaneous
reflection $(\tau,\mu)\leftrightarrow -(\tau,\mu)$, the propagator is
invariant. Furthermore, both the zero temperature and the finite
temperature propagators satisfy the equation
\begin{equation}
\left((\partial_{\tau} - \mu)^{2} - E^{2}\right)\Delta^{(T,\mu)}
(\tau,E) = - \delta (\tau).
\end{equation}
The finite temperature propagator, however, satisfies the periodicity
condition (\ref{kms}).

The form of the thermal propagator in the presence of a chemical
potential is rather complicated and it is not clear whether there will
be a factorization in this case. A little bit of analysis, however,
shows that the propagator can, in fact, be factorized as
\begin{eqnarray}
\Delta^{(T,\mu)} (\tau,E) & = &
\frac{1}{2E}\left[\theta(\tau)\left\{(1 + n_{-})e^{-(E-\mu)\tau} 
+ n_{+} e^{(E+\mu)\tau}\right\} + \theta (-\tau)\left\{n_{-} e^{-(E-\mu)\tau}
+ (1 + n_{+})e^{(E+\mu)\tau}\right\}\right]\nonumber\\
 & = & {\hat{\cal O}}^{(T,\mu)}(E,\partial_{\tau}) \Delta^{(T=0,\mu)}
 (\tau,E),\label{factorization1}
\end{eqnarray}
where
\begin{equation}
{\hat{\cal O}}^{(T,\mu)}(E,\partial_{\tau}) = \left[1 + \frac{n_{+} +
  n_{-}}{2} (1 - S(E))
  + \frac{n_{+}-n_{-}}{2} (1+S(E))
  \frac{1}{E}\left(\partial_{\tau}-\mu\right)\right].
\end{equation}
We note that the first two terms are similar to the ones in the
thermal operator of the earlier section while the last group of terms in the above
relation is new and vanishes when 
$\mu=0$. It is this term that reflects the asymmetry in the
chemical potential for particles and anti-particles. This basic
thermal operator reduces to the one in (\ref{factorization})
and continues to be independent of the time coordinate. However, we
would like to point out a further simplification that takes place in
this case. 

Let us note from (\ref{zeroTmupropt}) and (\ref{Tmuprop}) that the
dependence of the chemical potential in the exponents of the
propagators completely factorizes,
\begin{eqnarray}
\Delta^{(T=0,\mu)} (\tau,E) & = & e^{\mu\tau} \Delta^{(T=0=\mu)}
(\tau,E),\nonumber\\
\Delta^{(T,\mu)} (\tau,E) & = & e^{\mu\tau}
\frac{1}{2E}\left[\theta(\tau)\left\{(1 + n_{-})e^{-E\tau} 
+ n_{+} e^{E\tau}\right\} + \theta (-\tau)\left\{n_{-} e^{-E\tau} + (1 +
n_{+})e^{E\tau}\right\}\right],
\end{eqnarray}
where $\Delta^{(T=0=\mu)} (\tau,E)$ is given in (\ref{zeroTpropt}). 
Since $e^{\mu\tau}$ factors out in both zero as well as finite
temperature propagators, we can write a simpler factorization for the thermal
propagator as
\begin{equation}
\Delta^{(T,\mu)} (\tau,E) = e^{\mu\tau} {{\cal
O}}^{(T,\mu)}(E,\partial_{\tau}) \Delta^{(T=0=\mu)} (\tau,E),
\end{equation}
where
\begin{equation}
{\cal O}^{(T,\mu)}(E,\partial_{\tau}) = \left[1 + \frac{n_{+} +
  n_{-}}{2} (1 - S(E)) + \frac{n_{+}-n_{-}}{2} (1+S(E))
  \frac{1}{E} \partial_{\tau}\right].\label{factorization2}
\end{equation}
Furthermore, let us note that in any 1PI graph involving closed loops,
the overall factor $e^{\mu\tau}$ would cancel and hence can be
ignored. (There are various ways
of seeing this. Since the time comes back to itself in a closed loop,
this factor reduces to identity. In terms of electrostatic analogy, if
the particle comes back to the starting point, there is no change in
voltage. Such a simplification, however, would not take place in a tree level
graph which is not 1PI.) As we have argued earlier, the thermal operator
representation of a graph is  a 
reflection of the factorization of the thermal propagator and so given
the factorization in (\ref{factorization1}) it would seem that we can
write a simple thermal
operator representation for any finite temperature graph even in the
presence of a chemical potential. In general, however, because of the
time derivative terms in the basic thermal operator, this
factorization is not as simple as in the case without a chemical
potential. In the case of graphs, where every propagator is connected
to an external time coordinate, the thermal operator representation
takes a simple form for any 1PI graph involving closed loops,
\begin{equation}
\int \prod_{i=1}^{I} \frac{\mathrm{d}^{3}k_{i}}{(2\pi)^{3}}
\prod_{v=1}^{V}\delta_{v}^{(3)}(k,p)
\gamma_{N}^{(T,\mu)} = \int \prod_{i=1}^{I}
\frac{\mathrm{d}^{3}k_{i}}{(2\pi)^{3}}
\prod_{v=1}^{V}\delta_{v}^{(3)}(k,p)  {\cal O}^{(T,\mu)} 
\gamma_{N}^{(T=0=\mu)},
\end{equation}
where
\begin{equation}
{\cal O}^{(T,\mu)} = \prod_{i=1}^{I} {\cal O}^{(T,\mu)}
(E_{i},\pm \partial_{\tau_{\alpha}}),
\end{equation}
where $\tau_{\alpha}$ is the external time coordinate to which the
propagator with energy $E_{i}$ is connected ($\pm$ in the 
time derivative represents the phase that may arise in changing the
argument to the external time coordinate). This is almost like the case when
there is no chemical potential.

\begin{figure*}[ht!]
\begin{center}
\includegraphics[scale=0.5]{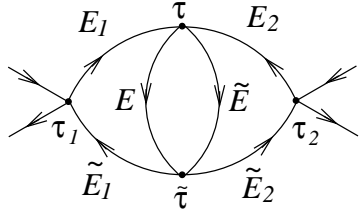}
\end{center}
\caption{A three-loop vertex correction diagram in the 
$(\phi^{*}\phi)^2$ theory with two internal time coordinates.}
\label{fig5}
\end{figure*}

However, if there are
propagators in a diagram which are not connected to an external time,
it is not clear {\em a priori} whether a thermal operator
representation can be written for such a graph. The difficulty arises
because if both the time coordinates associated with a propagator are
internal times, it would seem that the basic thermal operator
(\ref{factorization2}) for a propagator cannot be taken out of the
time integration and, consequently, it is not clear whether a thermal
operator representation of the graph can result. That such a
representation, be it
nontrivial, may arise can be seen from a simple nontrivial graph
like the one shown in figure \ref{fig5}.

Let us consider the vertex correction graph at three loops in the
complex $\phi^{4}$ theory. The graph involves two internal times
$\tau,\tilde{\tau}$ that need to be integrated over and there are two
propagators with time coordinates that are completely internal. We
consider the graph with the charge flows as shown in the figure. In
this case, we note that we can write
\begin{eqnarray}
\gamma_{4}^{(T,\mu)} & = & \int \mathrm{d}\tau \mathrm{d}\tilde{\tau}\
\Delta^{(T,\mu)} (\tau_{1}-\tau,E_{1})\Delta^{(T,\mu)}
(\tau_{2}-\tau,E_{2}) \Delta^{(T,\mu)}
(\tilde{\tau}-\tau_{1},\tilde{E}_{1}) \Delta^{(T,\mu)}
(\tilde{\tau}-\tau_{2},\tilde{E}_{2})\nonumber\\
 &  & \qquad \times \Delta^{(T,\mu)}
(\tau-\tilde{\tau},E) \Delta^{(T,\mu)}
(\tau-\tilde{\tau},\tilde{E})\nonumber\\
 & = & \prod_{i=1}^{2}{\cal O}^{(T,\mu)} (E_{i},\partial_{\tau_{i}}^{(E_{i})})
 {\cal O}^{(T,\mu)} (\tilde{E}_{i}, -\partial_{\tau_{i}}^{(\tilde{E}_{i})}) \int
 \mathrm{d}\tau \mathrm{d}\tilde{\tau}\ \Delta^{(T=0=\mu)}
 (\tau_{1}-\tau,E_{1})\Delta^{(T=0=\mu)}
 (\tau_{2}-\tau,E_{2})\nonumber\\
&  & \qquad \times\Delta^{(T=0=\mu)}
 (\tilde{\tau}-\tau_{1},\tilde{E}_{1}) \Delta^{(T=0=\mu)}
 (\tilde{\tau}-\tau_{2},\tilde{E}_{2})
 \Delta^{(T,\mu)} (\tau-\tilde{\tau},E)
 \Delta^{(T,\mu)} (\tau-\tilde{\tau},\tilde{E}),
\end{eqnarray}
where the superscript on $\partial_{\tau_{i}}$ is to specify on which
propagator the time derivative is going to act. To see how the time
derivatives in the last two propagators can be taken out of the
integral, let us note that we can write
\begin{equation}
\Delta^{(T,\mu)} (\tau-\tilde{\tau},E) = \left(X(E) + Y
(E)\frac{1}{E}\partial_{\tau}\right)\Delta^{(T=0=\mu)}
(\tau-\tilde{\tau},E),
\end{equation}
where we have identified for simplicity
\begin{equation}
X (E) = 1 + \frac{n_{+}+n_{-}}{2} (1-S(E)),\quad Y (E) =
\frac{n_{+}-n_{-}}{2} (1+S(E)).
\end{equation}
Using this as well as using the identities
\begin{eqnarray}
\Delta^{(T=0=\mu)} (\tau-\tilde{\tau},E)\Delta^{(T=0=\mu)}
(\tau-\tilde{\tau},\tilde{E}) & = &
\frac{2(E+\tilde{E})}{(2E)(2\tilde{E})}\ \Delta^{(T=0=\mu)}
(\tau-\tilde{\tau},E+\tilde{E}) \nonumber\\
 & = & \frac{1}{E}\partial_{\tau}\Delta^{(T=0=\mu)} (\tau-\tilde{\tau},E)\
 \frac{1}{\tilde{E}} \partial_{\tau}\Delta^{(T=0=\mu)}
 (\tau-\tilde{\tau},\tilde{E}),\nonumber\\
\frac{1}{E}\partial_{\tau}\Delta^{(T=0=\mu)} (\tau-\tilde{\tau},E)\
\Delta^{(T=0=\mu)} (\tau-\tilde{\tau},\tilde{E}) & = &
\frac{1}{E+\tilde{E}}\ \partial_{\tau}\left(\Delta^{(T=0=\mu)}
(\tau-\tilde{\tau},E) \Delta^{(T=0=\mu)}
(\tau-\tilde{\tau},\tilde{E})\right), 
\end{eqnarray}
we can write
\begin{eqnarray}
&  & \Delta^{(T,\mu)} (\tau-\tilde{\tau},E) \Delta^{(T,\mu)}
(\tau-\tilde{\tau},\tilde{E})\nonumber\\
&  &\quad = \left[X(E)X(\tilde{E}) + Y(E)
  Y(\tilde{E})\right] \Delta^{(T=0=\mu)}
(\tau-\tilde{\tau},E)\Delta^{(T=0=\mu)}
(\tau-\tilde{\tau},\tilde{E})\nonumber\\
 &  & \qquad + \left[X(E)Y(\tilde{E}) +
   X(\tilde{E})Y(E)\right]\frac{1}{E+\tilde{E}}
 \partial_{\tau}\left(\Delta^{(T=0=\mu)} (\tau-\tilde{\tau},E)
   \Delta^{(T=0=\mu)} (\tau-\tilde{\tau},\tilde{E})\right).
\end{eqnarray}
The $\tau$ derivative can now be integrated by parts inside the
integral and put on the other propagators where the argument of the
derivative can be changed to an external time coordinate and can be
taken outside the integral. This shows that although naively we will
not expect a factorization of the graph in figure \ref{fig5} where
there are propagators
containing only internal time coordinates, a thermal operator
representation does exist. However, it is not as simple as in the case
without a chemical potential and furthermore a closed form of the
thermal operator can only be determined graph by graph. 
In view of the above analysis, it seems plausible that such a
non-trivial factorization may also hold for general diagrams.

\section{Closed Time Path Formalism}

As we have emphasized several times earlier, the thermal operator
representation follows much more directly in the real time formalism
of closed time 
path \cite{das:book97,Schwinger:1961qe,Bakshi:1962dvKeldysh:1964ud}.
Let us recall that in the closed time path
formalism, the theory is defined in Minkowski space where time is a
continuous real variable defined over $(-\infty,\infty)$, unlike in
the imaginary time formalism. Of course,
the price one has to pay is to double the number of degrees of
freedom, for every field in our theory (we denote the real scalar
field of our theory by $\phi_{+}$) we add another field of the same
kind $\phi_{-}$. (We refer the readers 
to \cite{das:book97}
for details.) As a
result, the propagator acquires a $2\times 2$ matrix structure and in
the momentum space has the form
\begin{equation}
\Delta^{(T)} (p) = \left(\begin{array}{cc}
\Delta_{++}^{(T)} (p) & \Delta_{+-}^{(T)} (p)\\
\noalign{\vskip 2pt}%
\Delta_{-+}^{(T)} (p) & \Delta_{--}^{(T)} (p)
\end{array}\right),
\end{equation}
where, for a massive real scalar field, we have
\begin{eqnarray}
\Delta_{++}^{(T)} (p) & = & \lim_{\epsilon\rightarrow 0}\
\frac{i}{p^{2}-m^{2} + i\epsilon} + 2\pi
n(|p_{0}|) \delta (p^{2}-m^{2}),\nonumber\\
\Delta_{+-}^{(T)} (p) & = & 2\pi \left(\theta(-p_{0}) +
n(|p_{0}|)\right) \delta (p^{2}-m^{2}),\nonumber\\
\Delta_{-+}^{(T)} (p) & = & 2\pi \left(\theta(p_{0}) +
  n(|p_{0}|)\right)\delta (p^{2}-m^{2}),\nonumber\\
\Delta_{--}^{(T)} (p) & = & \lim_{\epsilon\rightarrow 0}\  -
\frac{i}{p^{2}-m^{2}-i\epsilon} + 2\pi n(|p_{0}|) \delta
(p^{2}-m^{2}),
\end{eqnarray}
with $n(|p_{0}|)$ denoting the bosonic distribution function.
The components at zero temperature follow from this to be
\begin{eqnarray}
\Delta_{++}^{(T=0)} (p) & = & \lim_{\epsilon\rightarrow 0}\
\frac{i}{p^{2}-m^{2}+i\epsilon},\nonumber\\
\Delta_{+-}^{(T=0)} (p) & = & 2\pi \theta (-p_{0}) \delta
(p^{2}-m^{2}),\nonumber\\
\Delta_{-+}^{(T=0)} (p) & = & 2\pi \theta (p_{0}) \delta
(p^{2}-m^{2}),\nonumber\\
\Delta_{--}^{(T=0)} (p) & = & \lim_{\epsilon\rightarrow 0}\ -
\frac{i}{p^{2}-m^{2}-i\epsilon}.
\end{eqnarray}

These are Minkowski space propagators and the $i\epsilon$ prescription
in the diagonal elements specifies the choice of the contour in the
complex energy plane (the two diagonal elements simply correspond to time
ordered and anti-time ordered propagators). Fourier transforming the
energy variable,  
\begin{equation}
\Delta (t,\vec{p}\ ) = \Delta (t,E) = \int \frac{\mathrm{d}p_{0}}{2\pi}\
e^{-ip_{0}t}\ \Delta (p),
\end{equation}
where, as before, $E = \sqrt{\vec{p}^{\ 2}+m^{2}}$, for the zero
temperature components we obtain
\begin{eqnarray}
\Delta_{++}^{(T=0)} (t,E) & = & \lim_{\epsilon\rightarrow 0}\
\frac{1}{2E}\left[\theta (t) e^{-i(E-i\epsilon)t} + \theta (-t)
e^{i(E-i\epsilon)t}\right],\nonumber\\
\Delta_{+-}^{(T=0)} (t,E) & = & \frac{1}{2E}\ e^{iEt},\nonumber\\
\Delta_{-+}^{(T=0)} (t,E) & = & \frac{1}{2E}\ e^{-iEt},\nonumber\\
\Delta_{--}^{(T=0)} (t,E) & = & \lim_{\epsilon\rightarrow 0}\
\frac{1}{2E}\left[\theta (t) e^{i(E+i\epsilon)t} + \theta (-t)
e^{-i(E+i\epsilon)t}\right]. \label{realzeroTprop}
\end{eqnarray}
Similarly, the Fourier transform of the finite temperature propagator
yields the components to be
\begin{eqnarray}
\Delta_{++}^{(T)} (t,E) & = & \lim_{\epsilon\rightarrow 0}\
\frac{1}{2E}\left[\theta(t) e^{-i(E-i\epsilon)t} +
\theta(-t)e^{i(E-i\epsilon)t} + n(E)(e^{-iEt} +
e^{iEt})\right],\nonumber\\
\Delta_{+-}^{(T)} (t,E) & = & \frac{1}{2E}\left[n(E) e^{-iEt} + (1 +
n(E)) e^{iEt}\right],\nonumber\\
\Delta_{-+}^{(T)} (t,E) & = & \frac{1}{2E}\left[(1 + n(E)) e^{-iEt} +
n(E) e^{iEt}\right],\nonumber\\
\Delta_{--}^{(T)} (t,E) & = & \lim_{\epsilon\rightarrow 0}\
\frac{1}{2E}\left[\theta(t) e^{i(E+i\epsilon)t} + \theta (-t)
e^{-i(E+i\epsilon)t} + n(E) (e^{-iEt} +
e^{iEt})\right].\label{realTprop}
\end{eqnarray}
We have carefully kept the $i\epsilon$ terms in the exponent resulting
from the Feynman prescription which are essential for the convergence
of factors in any calculation.

Looking at the components of the propagators in the mixed space in
(\ref{realzeroTprop}) and (\ref{realTprop}) we see that there is
natural factorization of the components of the finite temperature
propagator so that we can write
\begin{eqnarray}
\Delta^{(T)} (t,E) & = &  \left(\begin{array}{cc}
\Delta_{++}^{(T)} (t,E) & \Delta_{+-}^{(T)} (t,E)\\
\noalign{\vskip 2pt}%
\Delta_{-+}^{(T)} (t,E) & \Delta_{--}^{(T)} (t,E)
\end{array}\right) = \left(1 + n(E) (1-S(E))L(\epsilon)\right)
\left(\begin{array}{cc}
\Delta_{++}^{(T=0)} (t,E) & \Delta_{+-}^{(T=0)} (t,E)\\
\noalign{\vskip 2pt}%
\Delta_{-+}^{(T=0)} (t,E) & \Delta_{-+}^{(T=0)} (t,E)
\end{array}\right)\nonumber\\
& = & \left(1 + n(E) (1-S(E))L(\epsilon)\right)
\Delta^{(T=0)} (t,E)
={\cal O}^{(T)}(E)\Delta^{(T=0)} (t,E).\label{realfactorization}
\end{eqnarray}
Here $L(\epsilon)$ is an operator that takes the limit
$\epsilon\rightarrow 0$ in the expression on which it acts. If there
is no $\epsilon$ dependence in the expression, the effect of this
operator is that of the identity operator. Thus, we see that there is
a very simple factorization of the thermal operator in the closed
time path formalism where each component of the matrix propagator
factorizes by the same factor which does not depend on time and is
reminiscent of
(\ref{factorization}) (we recall that there is no $\epsilon$
dependence in the imaginary time formalism). This is, however, not the
case in other real time descriptions such as thermofield dynamics as
we will discuss in the next section.

\subsection{General Proof}

Given the simple factorization (\ref{realfactorization}) of the finite
temperature matrix propagators, the thermal operator representation
for any $N$-point graph now follows immediately. Let us note that any
graph in the closed time path formalism is simply a product of
vertices (both ``$+$''and ``$-$''types) and the corresponding
propagators with integrations over internal time coordinates. Since
each component of the propagator has the same simple factorization
(\ref{realfactorization}), then it follows that for any $N$-point
graph (with only ``$+$'' vertices or ``$-$'' vertices or mixed
vertices) where there are no internal time coordinates,
\begin{equation}
\int \prod_{i=1}^{I} \frac{\mathrm{d}^{3}k_{i}}{(2\pi)^{3}}\
\prod_{v=1}^{V} {(2\pi)^{3}}  \delta_{v}^{(3)}(k,p)
\gamma_{N}^{(T)} = \int \prod_{i=1}^{I}
\frac{\mathrm{d}^{3}k_{i}}{(2\pi)^{3}}\ 
\prod_{v=1}^{V} {(2\pi)^{3}}  \delta_{v}^{(3)}(k,p)\,
{\cal O}^{(T)}\,
\gamma_{N}^{(T=0)},\label{realtor}
\end{equation}
where $\gamma_{N}^{(T=0)}$ represents the value of the corresponding
Minkowski space graph after the energy integrations have been carried
out and 
\begin{equation}
{{\cal O}}^{(T)} = \prod_{i=1}^{I} {{\cal O}}^{(T)}(E_{i}) =
\prod_{i=1}^{I} \left(1 + n_{i} (1-S_{i})L(\epsilon)\right).
\end{equation}
This is exactly the same as in the imaginary time formalism. When
there are internal time coordinates that need to be integrated over,
however, the proof of the thermal operator representation in the
closed time path formalism is much simpler. In fact, note that here
time is a real variable defined over the entire real axis independent
of whether we are at zero temperature or at finite temperature. This
is the difference from the imaginary time formalism. As a result, the
thermal operator representation (\ref{realtor}) continues to hold even
when  there
are internal time coordinates that need to be integrated
over. (Namely, in this case, we do not have to extend the range of
integration and thereby avoid the complicated proof of equivalence in
extending the range of integration as is needed in the imaginary
time.) We find this proof of the thermal operator representation by
far the simplest. Furthermore, this result holds for any $N$ point
amplitude including the case when $N=0$. (Namely, we have made no
assumption about a graph having an external leg.) Therefore, the
thermal operator representation clearly holds even for graphs without
any external legs (such as pressure). We have not been able to
show this directly in the imaginary time formalism although we have
argued (based on our result of the closed time path formalism) and
shown in simple examples that it must be true. 

It is worth pointing out some features of the action of
the thermal operator at this point. First of all, let us consider a
multi-loop graph with all external vertices only of ``$+$''type. At
finite temperature, we know that there can be internal vertices of
``$-$''type. At zero temperature, however, the ``$-$''type
internal vertices give vanishing contribution\footnote{We would
like to thank P. Bedaque for a discussion on this point.}. There are
several ways of seeing this. The simplest reason is probably the most
physical, namely, at zero temperature any amplitude is simply given by
the original theory (one does not need a doubling at zero
temperature). Thus, applying the thermal operator representation to
such a graph would imply
\begin{equation}
\gamma_{+++\cdots}^{(T)} = {{\cal O}}^{(T)} \gamma_{+++\cdots}^{(T=0)}.
\end{equation}
Since the zero temperature graphs would not involve any ``$-$''type
intermediate vertices, this would imply that the finite temperature
result for such an amplitude can be obtained only from graphs
involving ``$+$''type vertices. This is, however, in contradiction
with the known fact that we need a doubling of the degrees of freedom
at finite temperature. The resolution of this puzzle is interesting,
which also clarifies
the action of the thermal operator in the following way. Namely, even
when a zero temperature graph vanishes, one should not set it to zero
before applying the thermal operator to it. Only after the thermal
operator has been applied can the relevant terms be set to zero. This
is also how new channels of 
reaction \cite{Weldon:1982aq}
at finite temperature can be seen
to arise in the thermal operator representation.

\begin{figure*}[ht!]
\begin{center}
\includegraphics[scale=0.5]{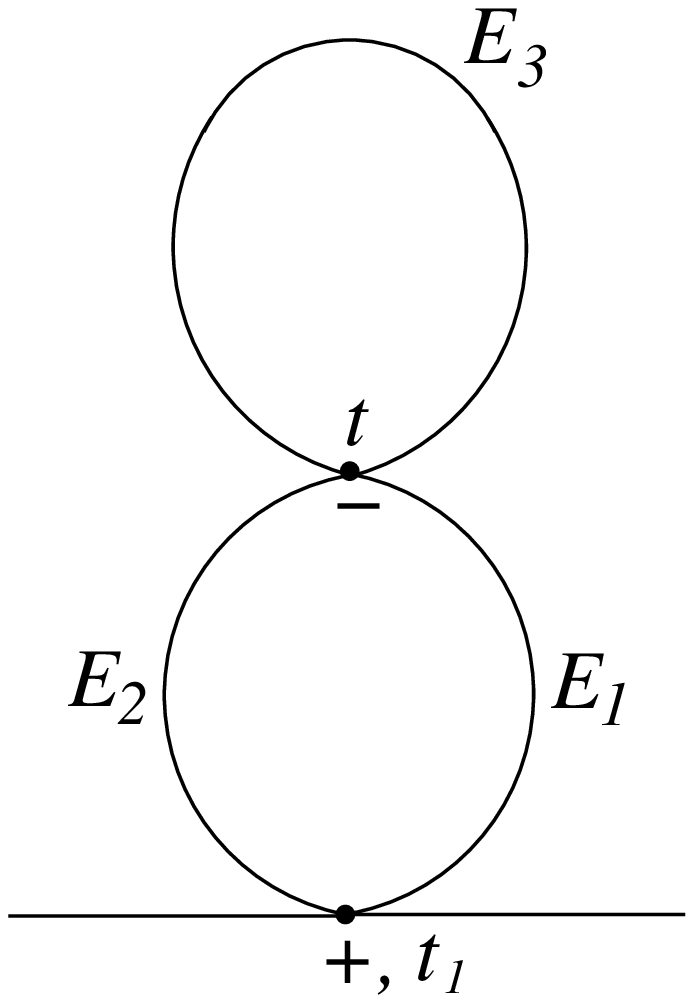}
\end{center}
\caption{A two loop self-energy diagram in the $\phi^4$ theory with an internal
 ``$-$'' vertex.}
\label{fig6}  
\end{figure*}

Let us illustrate this with the two loop example in the
$\phi^{4}$ field theory shown in figure \ref{fig6}.
At zero temperature, this graph leads to (we factor out
overall factors involving coupling and symmetry factors)
\begin{eqnarray}
\gamma_{+}^{(T=0)} & = & - \int_{-\infty}^{\infty} \mathrm{d}t\
\Delta_{+-}^{(T=0)}(t_{1}-t,E_{1}) \Delta_{+-}^{(T=0)} (t_{1}-t,E_{2})
\Delta_{--}^{(T=0)} (0,E_{3})\nonumber\\
 & = & - \frac{1}{(2E_{1})(2E_{2})(2E_{3})}\ 2\pi \delta
 (E_{1}+E_{2}).
\end{eqnarray}
Since both $E_{1},E_{2}$ are positive, this is clearly zero and is
consistent with our observation that at zero temperature, graphs with
internal ``$-$''type vertices do not contribute. At finite
temperature, however, this graph does contribute and has the value
\begin{eqnarray}
\gamma_{+}^{(T)} & = & - \int_{-\infty}^{\infty} \mathrm{d}t\
\Delta_{+-}^{(T)}(t_{1}-t,E_{1})  \Delta_{+-}^{(T)} (t_{1}-t,E_{2})
\Delta_{--}^{(T)} (0,E_{3})\nonumber\\
 & = & - \frac{1}{(2E_{1})(2E_{2})(2E_{3})}\
 2n_{1}(1+n_{1})(1+2n_{3})\times 2\pi \delta (E_{1}-E_{2}).\label{ft}
\end{eqnarray}
If we naively set the zero temperature graph to zero, then clearly
there is a contradiction. On the other hand, if we apply the thermal
operator to the zero temperature result, we obtain
\begin{eqnarray}
&  & \left(1+n_{1}(1-S_{1})\right)\left(1 +
  n_{2}(1-S_{2})\right)\left(1 + n_{3}(1-S_{3})\right)
(-1)\frac{1}{(2E_{1})(2E_{2})(2E_{3})}\ 2\pi\delta
(E_{1}+E_{2})\nonumber\\
& =& - \frac{2\pi}{(2E_{1})(2E_{2})(2E_{3})}\
(1+2n_{3})\left[\left\{(1+n_{1})(1+n_{2}) +
    n_{1}n_{2}\right\}\delta(E_{1}+E_{2}) + \left\{(1+n_{1})n_{2} +
    n_{1}(1+n_{2})\right\} \delta (E_{1}-E_{2})\right]. \nonumber\\
 &  & 
\end{eqnarray}
If we now use the fact that both $E_{1},E_{2}$ are positive to set the
first group of terms to zero and use the properties of the second
delta function, we obtain the finite temperature result in (\ref{ft}).
This simple example is quite illustrative in understanding the action of the
thermal operator.

\subsection{Chemical Potential}

The simplicity of the closed time path formalism continues to hold
even in the presence of a chemical potential as we will show without
going into too many details. Let us recall that the free Lagrangian
density for a complex scalar field with a chemical potential has the
form given in (\ref{lagrangian})
\begin{equation}
{\cal L} = \left(\partial_{t}+ i\mu\right)\phi^{*}
\left(\partial_{t}-i\mu\right)\phi - \vec{\nabla}\phi^{*}\cdot
\vec{\nabla}\phi - m^{2}\phi^{*}\phi 
\end{equation}
In the closed time path formalism, we have to double the degrees of
freedom and so introducing the doubled degrees of freedom
$(\phi_{-},\phi^{*}_{-})$ (we label the original fields as
$(\phi_{+},\phi^{*}_{+})$), we note that the propagator acquires a
$2\times 2$ matrix structure (and carries a direction from $\phi$ to
$\phi^{*}$)
\begin{equation}
\Delta^{(T,\mu)} (t,E) = \left(\begin{array}{cc}
\Delta_{++}^{(T,\mu)} (t,E) & \Delta_{+-}^{(T,\mu)} (t,E)\\
\noalign{\vskip 2pt}%
\Delta_{-+}^{(T,\mu)} (t,E) & \Delta_{--}^{(T,\mu)} (t,E)
\end{array}\right),
\end{equation}
with
\begin{eqnarray}
\Delta_{++}^{(T,\mu)} (t,E) & = & \lim_{\epsilon\rightarrow 0}
\frac{1}{2E}\left[\theta(t) e^{-i(E-\mu-i\epsilon)t} + \theta(-t)
  e^{i(E+\mu-i\epsilon)t} + n_{-}e^{-i(E-\mu)t} +
  n_{+}e^{i(E+\mu)t}\right]\nonumber\\
 & = & \lim_{\epsilon\rightarrow 0}\ \frac{e^{i\mu
     t}}{2E}\left[\theta(t) e^{-i(E-i\epsilon)t} + \theta(-t)
  e^{i(E-i\epsilon)t} + n_{-}e^{-iE t} +
  n_{+}e^{iE t}\right],\nonumber\\
\Delta_{+-}^{(T,\mu)} (t,E) & = & \frac{1}{2E}\left[n_{-}e^{-i(E-\mu)t} +
  (1+n_{+}) e^{i(E+\mu)t}\right]\nonumber\\
 & = & \frac{e^{i\mu t}}{2E}\left[n_{-}e^{-iE t} +
  (1+n_{+}) e^{iE t}\right],\nonumber\\
\Delta_{-+}^{(T,\mu)} (t,E) & = &
\frac{1}{2E}\left[(1+n_{-})e^{-i(E-\mu)t} + n_{+}
  e^{i(E+\mu)t}\right]\nonumber\\
 & = & \frac{e^{i\mu t}}{2E}\left[(1+n_{-})e^{-iE t} + n_{+}
  e^{iE t}\right],\nonumber\\ 
\Delta_{--}^{(T,\mu)} (t,E) & = & \lim_{\epsilon\rightarrow 0}
\frac{1}{2E}\left[\theta(t) e^{i(E+\mu+i\epsilon)t} + \theta (-t)
  e^{-i(E-\mu+i\epsilon)t} + n_{-}e^{-i(E-\mu)t} + n_{+}
  e^{i(E+\mu)t}\right]\nonumber\\
 & = & \lim_{\epsilon\rightarrow 0}
\frac{e^{i\mu t}}{2E}\left[\theta(t) e^{i(E+i\epsilon)t} + \theta (-t)
  e^{-i(E+i\epsilon)t} + n_{-}e^{-iE t} + n_{+}
  e^{iE t}\right] .\label{realTmuprop}
\end{eqnarray}
Here $n_{\pm}$ are the distribution functions introduced earlier in
(\ref{distribution}). 

Once again, it is obvious that component by component, we have a
factorization of the thermal operator as
\begin{equation}
\Delta_{ij}^{(T,\mu)} (t,E) = e^{i\mu t} {\cal O}^{(T,\mu)}(E,\partial_{t})
\Delta_{ij}^{(T=0=\mu)} (t,E),
\end{equation}
where $i,j=\pm$ and
\begin{equation}
{\cal O}^{(T,\mu)}(E,\partial_{t}) = \left[1 +
  \frac{n_{+}+n_{-}}{2}(1-S(E))L(\epsilon) 
  - \frac{n_{+}-n_{-}}{2} (1+S(E))L(\epsilon) \frac{i}{E}
  \partial_{t}\right]
\end{equation}
This simple factorization of every component of the thermal propagator
as well as the fact that in the closed time path formalism, the range
of integration over internal time coordinates is the same at zero as
well as at finite temperatures immediately leads to the thermal
operator representation for any 1PI graph. However, as discussed in
the case of the imaginary time formalism, the thermal operator
representation in the presence of a chemical potential is nontrivial
and more involved because of the presence of the time derivative terms
in ${\cal O}^{(T,\mu)} (E,\partial_{t})$.

\section{General Real Time Contour}

\begin{figure*}[ht!]
\begin{center}
\includegraphics[scale=0.5]{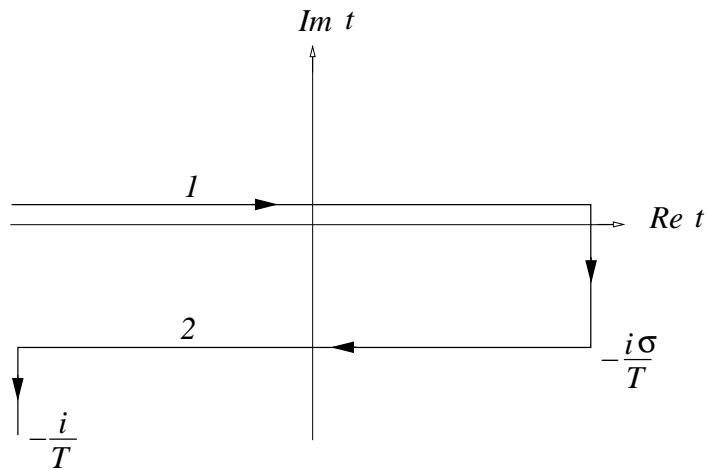}
\end{center}
\caption{General time contour in the complex $t$-plane.}
\label{fig7}  
\end{figure*}

There are two commonly used real time descriptions of finite
temperature field theory. In the last section, we have already
discussed one of them, namely, the closed time path 
formalism \cite{das:book97,Schwinger:1961qe,Bakshi:1962dvKeldysh:1964ud}
which is
quite useful in calculating various quantities both in thermal
equilibrium as well as out of thermal equilibrium. The other commonly
used real time formalism goes under the name of thermofield 
dynamics \cite{das:book97,umezawa:1982nv}
which is inherently an operator formalism and is quite useful in
understanding various operatorial issues such as the nature of the
thermal vacuum and the thermal Hilbert space etc for equilibrium
systems. It can also be used for calculations in thermal equilibrium
although its main power lies in understanding operatorial issues. Both
these formalisms correspond to specific time paths in the complex
$t$-plane. In general, one can define a thermal field theory with a
time contour in the complex $t$-plane of the form 
shown in figure \ref{fig7} \footnote{See for example pg. 60 in
  \cite{das:book97} under the scaling $\sigma\rightarrow
  \frac{\sigma}{T}$.}.
Here $\sigma$ is a constant and has the value in the range $0\leq
\sigma\leq 1$. When $\sigma = 0$ the description of the
thermal field theory coincides with the closed time path formalism
while for $\sigma = \frac{1}{2}$, the description corresponds to
thermofield dynamics. In this section, we will show that a simple
factorization of the thermal operator and, therefore, of a thermal graph
takes place only for the cases when $\sigma =0, 1$. In the
case of closed time path corresponding $\sigma=0$, we have already
seen this. Here we will study the behavior of the thermal propagator
for a general time path in the complex $t$-plane.

Let us consider a theory describing a real scalar field. The doubling
at finite temperature simply corresponds to introducing the fields
$(\phi_{1},\phi_{2})$ on the two paths labelled $1,2$ (the original
field that we start out with is considered to be $\phi_{1}$). For a
general time contour as shown in Fig.4, it can be shown that

\begin{equation}
\Delta^{(T)} (p) = \left(\begin{array}{cc}
\Delta_{11}^{(T)} (p) & \Delta_{12}^{(T)} (p)\\ \noalign{\vskip 2pt}%
\Delta_{21}^{(T)} (p) & \Delta_{22}^{(T)} (p)
\end{array}\right),
\end{equation}
where
\begin{eqnarray}
\Delta_{11}^{(T)} (p) & = & \Delta_{++}^{(T)} (p),\nonumber\\
\Delta_{12}^{(T)} (p) & = & e^{\frac{\sigma p_{0}}{T}}
\Delta_{+-}^{(T)} (p),\nonumber\\ 
\Delta_{21}^{(T)} (p) & = & e^{-\frac{\sigma p_{0}}{T}} \Delta_{-+}^{(T)}
(p),\nonumber\\
\Delta_{22}^{(T)} (p) & = & \Delta_{--}^{(T)} (p),
\end{eqnarray}
where $\Delta_{ij}^{(T)} (p), i,j=\pm$ correspond to the propagators
in the closed time path formalism that we have already discussed in
the earlier section. Fourier transforming these components of the
propagator in the energy variable, we obtain the mixed space
propagator for the general contour to be
\begin{eqnarray}
\Delta_{11}^{(T)} (t,E) & = & \Delta_{++}^{(T)} (t,E),\nonumber\\
\Delta_{12}^{(T)} (t,E) & = & \Delta_{+-}^{(T)} (t+i\frac{\sigma}{T},E) = 
\frac{1}{\sinh\frac{E}{T}} \left[\sinh\left((1-\sigma)\frac{E}{T}\right)\
  \Delta_{+-}^{(T)} (t,E)
  + \sinh\left(\frac{\sigma E}{T}\right)\  \Delta_{-+}^{(T)}
  (t,E)\right],\nonumber\\
\Delta_{21}^{(T)} (t,E) & = & \Delta_{-+}^{(T)} (t-i\frac{\sigma}{T},E) = 
\frac{1}{\sinh\frac{E}{T}} \left[\sinh\left(\frac{\sigma E}{T}\right)\
  \Delta_{+-}^{(T)} (t,E) + \sinh\left(
  (1-\sigma) \frac{E}{T}\right)\  \Delta_{-+}^{(T)} (t,E)\right],\nonumber\\
\Delta_{22}^{(T)} (t,E) & = & \Delta_{--}^{(T)} (t,E).\label{generalprop}
\end{eqnarray}
We note from (\ref{generalprop}) that in the limit $T\rightarrow 0$,
\begin{eqnarray}
\Delta_{12}^{(T)} (t,E) &\rightarrow& e^{-\frac{\sigma E}{T}}\
\Delta_{+-}^{(T=0)} (t,E) + 
e^{(\sigma - 1)\frac{E}{T}}\ \Delta_{-+}^{(T=0)} (t,E)\nonumber\\
 & = & \delta_{\sigma,0}\ \Delta_{+-}^{(T=0)} (t,E) +
 \delta_{\sigma,1}\ \Delta_{-+}^{(T=0)} (t,E),\nonumber\\
\Delta_{21}^{(T)} (t,E) &\rightarrow& e^{(\sigma-1)\frac{E}{T}}\
\Delta_{+-}^{(T=0)} (t,E) + e^{-\frac{\sigma E}{T}}\
\Delta_{-+}^{(T=0)} (t,E)\nonumber\\
 & = & \delta_{\sigma,1}\ \Delta_{+-}^{(T=0)} (t,E) +
 \delta_{\sigma,0}\ \Delta_{-+}^{(T=0)} (t,E).
\end{eqnarray}
This shows that if $\sigma\neq 0,1$, the off-diagonal components of
the propagator vanish (at $T=0$) leading to a decoupling of the two fields at
zero temperature. When $\sigma=0,1$, the off-diagonal elements of the
components do not vanish at zero temperature, nonetheless there is
decoupling of the two degrees of freedom.

Since each component of the propagator in the closed time formalism
has a simple factorization given by (\ref{realfactorization}), it
follows from (\ref{generalprop}) that for a general time contour we
can write
\begin{equation}
\Delta ^{(T)} (t,E) = \left(1 +
  n(E)(1-S(E))L(\epsilon)\right)\tilde{\Delta} (t,E),
\end{equation}
where
\begin{eqnarray}
\tilde{\Delta}_{11} (t,E) & = & 
\Delta_{++}^{(T=0)} \nonumber\\
\tilde{\Delta}_{12} (t,E) & = & \frac{1}{\sinh\frac{E}{T}}
\left[\sinh\left((1-\sigma)\frac{E}{T}\right) \Delta_{+-}^{(T=0)}
  + \sinh\left(\frac{\sigma E}{T}\right)
  \Delta_{-+}^{(T=0)}\right]\nonumber\\
\tilde{\Delta}_{21} (t,E) & = & \frac{1}{\sinh\frac{E}{T}}
\left[\sinh\left(\frac{\sigma E}{T}\right) \Delta_{+-}^{(T=0)} + \sinh\left(
  (1-\sigma) \frac{E}{T}\right) \Delta_{-+}^{(T=0)}\right]\nonumber\\
\tilde{\Delta}_{22} (t,E) & = & \Delta_{--}^{(T=0)}.
\end{eqnarray}
It is clear from the above result that in the general case,
temperature cannot be completely factored out of the matrix in a
simple manner unless $\sigma = 0,1$. For the case of thermofield
dynamics where $\sigma = \frac{1}{2}$, it is easy to show that
\begin{equation}
\Delta^{(T)} (t,E) = \left(1 +
  n(E)(1-S(E))L(\epsilon)\right)\left(\begin{array}{cc}
1 & e^{-\frac{E}{2T}} L(\epsilon)\\
e^{-\frac{E}{2T}} L (\epsilon) & 1
\end{array}\right) \Delta^{(T=0)} (t,E),
\end{equation}
so that, in this case, the basic thermal operator takes a matrix
form. As a result, the thermal operator representation for any graph
in the formalism of a general contour (where $\sigma\neq 0,1$) is not
so simple as in the imaginary time formalism or the closed time path
formalism. However, it is worth noting that although in such cases
there will be no simple factorization at the level of
individual graphs, the simple factorization will occur for the
complete set of graphs associated with a given physical amplitude
(which follows from the fact that a physical amplitude is the same in
any formalism).

\section{Summary}

In this paper, we have systematically studied the interesting question
of thermal operator representation for Feynman graphs at finite
temperature. By working in a mixed space $(\tau,\vec{p}\ )$ (or
$(t,\vec{p}\ )$), we have given a simpler derivation of the thermal
operator representation in the imaginary time formalism. We have
traced the origin of such a simple relation to the fact that the
thermal propagator, in this space, has a basic factorization where the
basic thermal operator is independent of time. We have also
generalized the thermal operator representation to the case where
there is a nontrivial chemical potential. In this case, although the
thermal propagator also factorizes, the basic thermal operator
involves a time derivative which leads to a thermal operator
representation for any graph that is highly nontrivial.  We have tried
to study various properties of the thermal operator and have shown
that it is a projection operator which projects functions into the
space where the KMS periodicity condition is satisfied. We have also shown that
there is a simple thermal operator representation in the closed time
path formalism. The derivation, in this case, is even simpler than
that for the imaginary time formalism. For a general time contour
(including the one for thermofield dynamics), however, the thermal
operator representation is not so simple as in the imaginary time
formalism and the closed time path formalism.

\vskip .7cm

\noindent{\bf Acknowledgment}
\medskip

This work
was supported in part by the US DOE Grant number DE-FG 02-91ER40685,
by MCT/CNPq as well as by FAPESP, Brazil and by CONICYT, Chile under grant
Fondecyt 1030363 and 7040057 (Int. Coop.).


\end{document}